\documentclass[11pt]{article}
 \pdfoutput=1
 \usepackage{dsfont}
 \usepackage{bbm}
\usepackage[left=0.8in,right=0.8in,a4paper]{geometry}
\usepackage[pdfstartview=FitH,pdfpagemode=None]{hyperref}
\usepackage{color}
\usepackage{slashed}
\usepackage{amsmath}
\usepackage{amsfonts,bm}
\usepackage{amssymb}
\usepackage{authblk}
\usepackage{graphicx}
\usepackage{caption}
\usepackage{subcaption}
\usepackage{float}
\usepackage{tikz}
 \usetikzlibrary{snakes}
\usepackage{textcomp}
 \usepackage{stmaryrd}

\usepackage{siunitx}
\usepackage{appendix}
\usepackage{listings}
\newcommand{\n}{\nonumber}
\newcommand{\be}{\begin{equation}}
\newcommand{\ee}{\end{equation}}
\newcommand{\nin}{\noindent}

\def\bea{\begin{eqnarray}}
\def\eea{\end{eqnarray}}
\newcommand{\eq}[1]{(\ref{#1})}

\newcommand{\cn}{{\cal N}}

\newcommand{\co}{{\cal O}}

\newcommand{\cq}{{\cal Q}}

\newcommand{\nn}{\nonumber}

\def\eqa{\begin{eqnarray}}
\def\eqae{\end{eqnarray}}
\def\eq{\begin{equation}}
\def\eqe{\end{equation}}
\def\be{\begin{equation}}
\def\ee{\end{equation}}
\def\bea{\begin{eqnarray}}
\def\eea{\end{eqnarray}}
\def\ba{\begin{array}}
\def\ea{\end{array}}
\def\bd{\begin{displaymath}}
\def\ed{\end{displaymath}}

\def\>{\rangle}
\def\<{\langle}
\def\a{\alpha}
\def\b{\beta}

\def\del{\delta}
\def\e{\epsilon}
\def\f{\phi}
  
\def\g{\gamma}

\def\j{\psi}

\def\m{\mu}
\def\n{\nu}
\def\w{\omega}

\def\q{\theta}

\def\s{\sigma}
\def\t{\tau}

\def\z{\zeta}

\def\L{\Lambda}

\def\pa{\partial}
\begin{document}
 \usetikzlibrary{snakes,arrows,shapes,positioning,decorations}

\tikzstyle{mybox} = [draw=black, fill=blue!10, very thick,
    rectangle, rounded corners, inner sep=10pt, inner ysep=10pt]
\tikzstyle{fancytitle} =[fill=black, text=white]

\numberwithin{equation}{section}


\thispagestyle{empty}

\begin{titlepage}

\begin{center}

{\Large \bf
Supersymmetric Mixed Boundary Conditions in AdS$_2$
\\\vspace{0.3cm}
 and DCFT$_1$ Marginal Deformations}\\

\vspace{1.5cm}

{\large   Diego H. Correa, Victor I. Giraldo-Rivera  and Guillermo A. Silva{\let\thefootnote\relax\footnote{{correa@fisica.unlp.edu.ar, vigirald@gmail.com, silva@fisica.unlp.edu.ar}}}}\\

\vspace{1cm}

{\it Instituto de F\'isica La Plata - CONICET \&\\
Departamento de F\'\i sica,  Universidad Nacional de La Plata\\ C.C. 67, 1900,  La Plata, Argentina}

\vspace{14pt}

\end{center}
\begin{abstract}
We consider different supersymmetric mixed boundary conditions for scalar and fermionic fields in $AdS_2$, searching for the dual description of a family of interpolating Wilson Loops in ABJM theory.
The family, which interpolates
between the bosonic 1/6 BPS loop and the 1/2 BPS loop, can be thought of as an exact marginal deformation in a defect CFT$_1$. Confronting this property against  holographic  correlators and vacuum energy corrections singles out a particular boundary condition  which we propose as dual to the interpolating family of Wilson loops.

\end{abstract}

\end{titlepage}

\setcounter{page}{1} \renewcommand{\thefootnote}{\arabic{footnote}}
\setcounter{footnote}{0}
\newpage

\tableofcontents


\section{Introduction}


In the AdS/CFT context, Wilson loops have a dual description in terms of open strings  \cite{Rey:1998ik,Maldacena:1998im}. Wilson loop data such as  curve profile and coupling to scalar and fermion fields are encoded as boundary conditions for the string worldsheets. Classical worldsheets with prescribed boundary conditions give the leading contribution to the strong coupling expansion of the Wilson loop vacuum expectation value, subleading corrections are obtained from a semiclassical expansion around the classical solution.

For the  ${\cal N} =4$ super Yang-Mills case, the open strings propagate in $AdS_5\times S^5$ and, depending on whether one imposes Dirichlet or Neumann boundary conditions on the $S^5$ directions, one accounts for either  Maldacena-Wilson loops, coupled to scalars and gauge fields,  \cite{Drukker:1999zq} or  ordinary {\it non-supersymmetric} Wilson loops, coupled only to the gauge potential  \cite{Alday:2007he}. In particular, for a straight Wilson line, the same dual classical string embedding, with induced $AdS_2$ geometry,  solves the Dirichlet and Neumann boundary problems\footnote{This also happens for the circular Wilson loop.}. Concomitantly, computing the string fluctuations one finds massless scalar fields for the $S^5$ directions. Hence, the possibility of describing different Wilson loops with the same bulk object correlates to the mass of   fluctuating fields belonging to the Breitenlohner-Freedman (BF) window \cite{BF}. In fact, inside the BF window more general possibilities arise  from linear combinations of Dirichlet and Neumann boundary conditions. These generically break the scale symmetry, and, as shown in \cite{Witten:2001ua}, turn out to be dual to  relevant deformations of the CFT by double trace operators  (see also \cite{Klebanov:1999tb,GM,GK,Hartman:2006dy}, and \cite{Herzog:2019bom} for recent developments on marginal deformations). Open strings with mixed boundary conditions, dual to ${\cal N} =4$ super Yang-Mills Wilson loops,  have been studied in \cite{Polchinski:2011im}, with
the ordinary and supersymmetric Wilson loops corresponding to the UV and IR fixed points of a renormalization group flow.
More importantly for the present paper, it was  shown in \cite{Polchinski:2011im} that  supersymmetry was only preserved  for Dirichlet boundary conditions on the $S^5$.

Wilson loops in ${\cal N} =6$ super Chern-Simons-matter or ABJM theory are dual to open strings in $AdS_4\times \mathbb{CP}^3$, and the aim of the present article is to  consider mixed boundary conditions for the latter. We will find various interesting distinctions with respect to the  ${\cal N} =4$ super Yang-Mills (SYM) case. Initially two distinct {\it supersymmetric} Wilson loops were found in ABJM: the $1/2$ BPS coupled to gauge fields, scalars and fermions \cite{Drukker:2009hy} and the bosonic 1/6 BPS   coupled to gauge fields and scalars \cite{Drukker:2008zx,Chen:2008bp,Rey:2008bh} (see \cite{Drukker:2019bev} for a recent review on ABJM Wilson loops). The dual description of 1/2 BPS Wilson loops is given in terms of Dirichlet boundary conditions for all the $\mathbb{CP}^3$ directions.  On the other hand, the dual description of  bosonic 1/6 BPS Wilson loops was argued to be given in terms of a  delocalized string along a $\mathbb{CP}^1\subset\mathbb{CP}^3$  \cite{Drukker:2008zx}, which  was later interpreted in terms of Neumann boundary conditions for those directions \cite{Lewkowycz:2013laa}. We will show below that in  ABJM theory not only Dirichlet but also Neumann boundary conditions are consistent with supersymmetry. Moreover, we will find several supersymmetric mixed boundary conditions  that interpolate between Dirichlet and Neumann. After analyzing all of them, we will argue that the one called {\sf Type III} in this notes should account  for the supersymmetric family of  Wilson loops that interpolates between the bosonic 1/6 BPS and the 1/2 BPS,  recently found in \cite{Ouyang:2015iza,Ouyang:2015bmy}.

The paper is organized as follows:  In section \ref{mixingbc} we  construct different sets of supersymmetric mixed boundary conditions that interpolate between Dirichlet and Neumann, and count the preserved supersymmetries in each case. We also identify the boundary terms that enable to obtain them from a variational problem. In sections \ref{correlators} and \ref{vacuum} we compute holographic 2-point correlators and 1-loop corrections to the vacuum energy for the different sets of supersymmetric mixed boundary conditions. Finally, in section \ref{discu}, and based on the results obtained, we make our proposal for the dual to the supersymmetric family of  Wilson loops. We close with a couple of appendices reviewing our spinor conventions in 2d and the interpolating Wilson loops family.


\section{String Worldsheets and Mixed Boundary Conditions}
\label{mixingbc}


In this section we analyze different boundary conditions for open strings in $AdS_4\times \mathbb{CP}^3$. Eventually, we would like to  identify those that correspond to a family of Wilson loops which interpolate between the 1/2 BPS and the bosonic 1/6 BPS \cite{Ouyang:2015iza,Ouyang:2015bmy}. The particular family of Wilson loops we will identify is reviewed in appendix \ref{susyWL}.

We start with the Nambu-Goto action for strings
\be
S_{\sf NG}=\frac{L^2}{2\pi\alpha'}\int d^2\sigma\sqrt{|g|},\qquad g_{\alpha\beta}=G_{\mu\nu}(X)\partial_\alpha X^\mu\partial_\beta X^\nu\,,
\label{NG}
\ee
 propagating in $AdS_4\times \mathbb{CP}^3$ with metric
\be
ds^2= ds^2_{AdS_4}+4ds^2_{\mathbb{C P}^3}.
\ee
$AdS_4$ is foliated with $AdS _2$ slices as,
\bea
ds^2_{AdS_4}=du^2+\cosh^2\!u\;ds^2_{AdS_2}+\sinh^{2}\! u\;d\phi^2,
\eea
and the $\mathbb{CP}^3$ metric reads
\bea
&& ds^2_{\mathbb{C P}^3}=\frac{1}{4}\big[d\a^2+\cos^2\frac{\a}{2}(d\q_1^2+\sin^2\q_1 d\varphi^2_1)+\sin^2\frac{\a}{2}(d\q_2^2+\sin^2\q_2 d\varphi^2_2)\nonumber\\
&&\qquad\qquad\qquad\qquad\qquad\qquad +\sin^2\frac{\a}{2}\cos^2\frac{\a}{2}(d\xi+\cos\q_1 d\q_1-\cos\q_2 d\q_2)^2\big].
\eea
Classical string solutions anchored at the $AdS$ boundary provide the starting point for the strong coupling expansion of Wilson loops expectation value ($L^2/\alpha'\gg1$).

We will be interested in the open string worldsheet extending along the $AdS_2$ factor, with constant $\mathbb {CP}^3$ coordinates\footnote{Since $\mathbb {CP}^3$ is homogeneous one can fix an arbitrary point without loss of generality. The choice made in \eqref{C3} corresponds to the $SU(4)$ orientation chosen in \eqref{calM} for the Wilson loop definition.}
\be
u(\sigma)=\alpha(\sigma)= \theta_1(\sigma)=0.
\label{C3}
\ee
This gives rise to a solution of the string equations of motion. By choosing Poincar\'e coordinates the induced worldsheet metric results\footnote{Alternatively, we could choose global coordinates for $AdS_2$, doing so we describe the circular Wilson loop.}
\be
\label{eq:PoinPatch}
ds^2_{ind}=\frac{d\tau^2+dy^2}{y^2},
\ee
which describes a straight Wilson loop at the boundary.

It is important to stress now that the embedding \eqref{C3} gives a solution to two different boundary values problems for the coordinates spanning a $\mathbb {CP}^1\subset\mathbb {CP}^3$:
\begin{alignat}{3}
 &\textsf{Dirichlet:}&  \qquad \theta_1(\tau,0) &=0,  &\quad&\label{Dir}
 \\	
 &\textsf{Neumann:} & \qquad \partial_y\theta_1(\tau,0) &= 0,  &\quad  \partial_y\varphi_1(\tau,0) &=0.\label{Neu}
\end{alignat}

Fixing the string to a point inside  $\mathbb {CP}^3$  preserves an $SU(3)\subset SU(4)$
and 1/2 of the supersymmetry which suggests identifying the Dirichlet problem \eqref{Dir} with the 1/2 BPS straight Wilson line.  On the other hand, the Neumann problem leaves unspecified the values $\theta_0$ and $\varphi_0$ for $\theta_1(\tau,0)$ and $\varphi_1(\tau,0)$ respectively. In this last case, through a $\kappa$-symmetry analysis, one finds that the set of solutions for all possible values of $\theta_0$ and $\varphi_0$  share four common supersymmetries and preserves $SU(2)\times SU(2)\subset SU(4)$ $R$-symmetry group. Hence, a string worldsheet with Neumann boundary conditions on $\theta_1$ and $\varphi_1$  is associated with the bosonic  1/6 BPS Wilson line \cite{Drukker:2008zx}.


\subsection{Free Fields in $AdS_2$ and Supersymmetry}


In what follows, we  consider quadratic fluctuations of bosonic and fermionic degrees of freedom around the embedding \eqref{C3}. We analyze the supersymmetry of the action governing their dynamics by turning to Lorentzian signature.

The action for the bosonic fluctuations is found by expanding (\ref{NG}) up to quadratic order using
\be
X^\mu = X_{\sf clas}^\mu + \delta X^\mu,
\ee
where $X_{\sf clas}^\mu$ represents the embedding \eqref{C3}. In  static gauge the longitudinal fluctuations are set to zero and one is left with eight transverse scalars in an  $AdS_2$   geometry. The two transverse fluctuations along $AdS_4$ have mass $m_B^2=2$, while the remaining six scalars along
$\mathbb{C P}^3$ are massless $m_B^2=0$.
Fermionic fluctuations are obtained from the quadratic piece in the Green-Schwarz (GS) which in the present case comprises a 10-dimensional Majorana spinor \cite{cvet},\cite{Mart}. After performing the reduction to 2 dimensions,  one obtains six massive  spinors with  $|m_F| = 1$ and two massless ones with  $m_F=0$ again on $AdS_2$  \cite{KK,FLS}.

The presence of  massless fermions constitutes  the  crucial difference with respect to the  $AdS_5\times S^5$ case and will eventually become the reason why boundary conditions other than Dirichlet can be consistent with supersymmetry in ABJM models.  The set of  fluctuations can be packed into 4 complex scalars, one massive and three massless, and 4 Dirac fermions, three massive and one massless. We will generically refer to each of them as $\phi$ and $\psi$ respectively, and we will be particularly  interested in the massless  modes. For completeness we remind the reader that in $AdS_2$  alternative quantizations for scalars and fermions arise  for $-\frac14\le m^2_B\le\frac34$  and $| m_F|\le \frac12$ respectively (see \cite{BF,Amsel:2008iz}).

Consider now the following action for a complex scalar $\phi$ and a Dirac fermion $\psi$,
\bea
S_{\sf bulk}= \frac{1}{2}\int d^2x \sqrt{|h|}( h^{\a\b}\pa_\a\f^* \pa_\b \f -m_B^2\f^*\f+i \bar{\psi}\g^\a D_\a\psi-m_F\bar{\psi}\psi ),
\label{action}
\eea
with $h_{\a\b}$ the induced  $AdS_2$  worldsheet metric now in Lorentzian signature
\be
\label{eq:PoinPatch0}
ds^2=\frac{dt^2-dy^2}{y^2}.
\ee
Under the condition $m_B^2 = m_F^2 - m_F$,  action \eqref{action} becomes invariant under the following supersymmetry transformations  \cite{Sakai:1984vm},
\be
\delta \phi = \bar\varepsilon \psi,\qquad
\delta \psi = -\big(i\gamma^\a \partial_\a \phi + m_F \phi \big)\varepsilon,
\label{susytrans}
\ee
where $\varepsilon$ is a Dirac  Killing spinor of  $AdS_2$,
\bea
\label{eq:KSE}
D_\a  \varepsilon+ \frac{i}{2}\g_\a\varepsilon=0.
\eea
In Poincar\'e coordinates \eqref{eq:PoinPatch}, the solution to the  Killing spinor equation reads
\begin{align}
\varepsilon(t,y)= {y^{-1/2}}\xi(t) + y^{1/2} i \gamma_0\, \dot\xi (t)\qquad \text{with}\qquad i\gamma_1 \xi(t)= \xi(t)
\qquad \text{and}\qquad \xi(t)=\xi_0 + t\  \xi_1.
\label{KSS2}
\end{align}
$\xi_0$ and $\xi_1$ are  constant spinors parametrizing four real supercharges. Summarizing, a complex  scalar of mass $m_B^2 = m_F^2 - m_F$   and a Dirac fermion of mass $m_F$ comprise a ${\cal N}=2$ supermultiplet in $AdS_2$  (see app. \ref{sec:2dConventions} for details).

Massless scalar fields in $AdS_2$ admit standard and alternate quantizations (corresponding to Dirichlet and Neumann boundary conditions respectively). For the fluctuations of the string dual to Wilson loops in ${\cal N}=4$ SYM, all massless scalars   transform under supersymmetry into fermions having $|m_F |=1$, and these  lay outside the fermion BF window. Hence, as shown in \cite{Polchinski:2011im}, only Dirichlet boundary conditions are consistent with supersymmetry. As we show below, the appearance of an $AdS_2$ supermultiplet comprising massless scalars and fermions becomes the key point to enable more general types of supersymmetric boundary conditions.


\subsection{Supersymmetric Boundary Conditions for Massless Fields}
\label{22}


In what follows we present some admissible supersymmetric boundary conditions for a ${\cal N}=2$ supersymmetry multiplet consisting of massless bosonic and fermion fields in $AdS_2$. The behavior of the fields near the boundary is
\bea
\phi(t,y) &\!\!=\!\!&
\big(\alpha(t)+ \cdots\big) + y \big(\beta(t)+ \cdots\big)
\label{expabos}
\\
\psi(t,y) &\!\!=\!\!& y^{1/2}\left( \alpha^\psi(t)
 + y\, \gamma_5 \dot\alpha ^\psi(t)
 +\cdots\right)
+ y^{1/2} \left(\beta^\psi(t)
+ y\, \gamma_5 \dot\beta ^\psi(t)
 +\cdots\right)
 \label{expafer}
\eea
where the ellipsis stands for higher orders in $y$ and $\alpha^\psi,\beta^\psi  $ are eigenvectors of the projectors along the radial direction $P_\pm$ defined in \eqref{ppm}
\begin{equation}
P_- \alpha^\psi = \alpha^\psi,\qquad
P_+ \beta^\psi = \beta^\psi.
\end{equation}
Notice that $\alpha^\psi$ and $\beta^\psi$ in (\ref{expafer}) show the same asymptotic behaviour at the boundary. As well known, this follows from  massless fermions having a unique falloff at the boundary $\left.\Delta_\pm^{F}\right|_{m_F=0}=\frac12$.
Using these results in  combination with (\ref{susytrans}) and  the Killing spinor  \eqref{KSS2}  one finds
\begin{alignat}{2}
\delta\beta  &=  \bar \xi \gamma_5 \dot \beta^\psi + \dot{\bar \xi}  \gamma_5  \beta^\psi, &\qquad  \quad \delta\alpha &=   \bar \xi \alpha^\psi
\label{scalarbc}
\\
\delta\beta^\psi &= \beta  \xi, &
\delta\alpha^\psi  &=   \dot\alpha \gamma_5 \xi\,.
\label{fermionbc}
\end{alignat}

We say that a set of boundary conditions is consistent with supersymmetry if they remain invariant under some supersymmetry transformation. It is immediate from (\ref{scalarbc}) that imposing a Dirichlet boundary condition on the scalar field, {\it i.e.} $\alpha(t) =0$,  will be invariant under supersymmetry if we impose $\alpha^\psi(t) = 0$ on the fermion. Conversely, condition $\alpha(t) =0$ guarantees that the fermionic counterpart $\alpha^\psi(t) = 0$ remains invariant. Therefore, setting $\alpha=\alpha^\psi=0$ defines a supersymmetric boundary conditions which we call Dirichlet.  Alternatively, imposing a Neumann condition on the scalar field, {\it i.e.} $\beta(t) =0$, in combination with $\beta^\psi(t) =0$ on the fermion allows for a second supersymmetric boundary condition which we call  Neumann. These are standard results for massless fields.

Summarizing, Dirichlet and Neumann boundary conditions are consistent with supersymmetry for the case of massless multiplets in $AdS_2$:
\begin{alignat}{3}
 &\textsf{Dirichlet:}&  \qquad \alpha(t) &= 0,  &\alpha^\psi(t) &=  0.~~~~4~\text{supersymmetries}\label{Dirbc}
 \\	
 &\textsf{Neumann:} & \qquad \beta(t) &= 0,  &\quad  \beta^\psi(t) &=0.~~~~4~\text{supersymmetries}\label{Neubc}
\end{alignat}
These two sets of supersymmetric boundary conditions concern transformations between massless scalar and  fermion fields, and the number of free parameters in the Killing spinor (\ref{KSS2}) show that both types preserve four supersymmetries, hence 1/6 BPS. Additional supersymmetries could arise from the transformation of the massless scalar fields into massive fermions (see \cite{Sakai:1984vm} and footnote 7 in \cite{Drukker:2000ep}). Since these other fermions have mass $|m_F|=1$, outside the fermionic BF window, additional supersymmetries can only be found for Dirichlet boundary conditions. These account for the enhancement to 1/2 BPS in \eqref{Dir}.

In what follows, we shall construct different types of mixed boundary conditions which interpolate between Dirchlet and Neumann. These two cases provide the dual description of the limiting examples of the aforementioned family of Wilson loops in ABJM. Given the fact that whole family of Wilson loops is supersymmetric, we shall restrict our attention to boundary conditions interpolating between Dirichlet and Neumann preserving some supersymmetry.


\subsubsection{Standard Mixed Boundary Conditions}
\label{sec:StanBC}


We start by considering a linear combination of Dirichlet and Neumann boundary conditions for the scalar field,
\begin{equation}
\chi \alpha  - \beta = 0.
\label{standardmix}
\end{equation}
This type of boundary condition has been shown to describe CFTs double trace deformations in the large $N$ limit (see  \cite{Witten:2001ua,Hartman:2006dy}) and more recently related to interpolating Wilson loops in  $\cn=4$ SYM  \cite{Polchinski:2011im,Beccaria:2017rbe}. It is perhaps the simplest way to interpolate between Dirichlet and Neumann boundary conditions. However, it turns out to be not adequate to describe the ABJM interpolating Wilson loops family at hand.

The first problem is that \eqref{standardmix} demands a restriction on the Killing spinor, such that it leads to only  two preserved supersymmetries. Condition \eqref{standardmix} remains invariant under a supersymmetry transformation if we impose the following constraint on the spinor
\begin{equation}
 \chi  \alpha^\psi -\gamma_5 \dot\beta^\psi  =0\,.
 \label{FbcS}
\end{equation}
Fortunately, this condition remains invariant if \eqref{standardmix} holds, however, only for those transformations arising from the $t$-independent component, $\xi_0$, of $\xi(t)$. As a consequence, conditions \eqref{standardmix}-\eqref{FbcS} preserve half the number of supersymmetries present in \eqref{KSS2}.

Another reason for them not being adequate to describe the family of Wilson loops is the fact that the interpolating parameter $\chi$ is dimensionful. This implies that the scale invariance of the dual 1-dimensional defect theory will be broken. This  is a well known result that becomes evident upon computing  correlation functions \cite{Hartman:2006dy}.

Combining  \eqref{standardmix} and \eqref{FbcS} we find
\begin{equation}
 \textsf{Type I:} \qquad \chi \alpha  - \beta  = 0, \quad \chi  \alpha^\psi -\gamma_5 \dot\beta^\psi  =0, \qquad 2~\text{supersymmetries}
 \label{T1}
\end{equation}
Decomposing the fields into real and imaginary parts, $\alpha = \alpha_1 + i\alpha_2$ and $\beta = \beta_1 +i \beta_2$, conditions \eqref{T1} turn into\footnote{Had we replaced $\chi\mapsto i\chi$ in \eqref{T1}, the decomposition into real and imaginary components would give
 \begin{alignat}{2}
  \qquad \chi  {\alpha}_1-\beta_2 & = 0, &
 \chi  {\alpha}_2+\beta_1= 0&,
 \nonumber
   \\
  \qquad \chi \alpha_1^\psi - \gamma_5 \dot\beta_2^\psi
  & = 0, & \qquad
 \quad  \chi \alpha_2^\psi +\gamma_5 \dot\beta_1^\psi = 0 &.
 \nonumber
\end{alignat}
 However, these boundary conditions cannot be derived by the addition of a boundary term to the action.}
\begin{alignat}{2}
 \qquad \chi  {\alpha}_1-\beta_1  & = 0, &
 \chi  {\alpha}_2-\beta_2 = 0&,\label{MixB}
   \\
  \chi \alpha_1^\psi - \gamma_5 \dot\beta_1^\psi   & = 0, &
  \qquad  \chi \alpha_2^\psi -\gamma_5 \dot\beta_2^\psi = 0&.\label{Neubc2}
\end{alignat}

For completeness, we would like to mention a supersymmetric boundary condition that arises in the context of ${\cal N}=1$ supersymmetry in $AdS_2$.   If we consider two ${\cal N}=1$ Wess-Zumino\footnote{A ${\cal N}=1$ Wess-Zumino supermultiplet in $d=2$ comprises a real scalar and a Majorana fermion. The supersymmetry transformations are simply \eqref{susytrans} for each of the multiplets. In terms of the asymptotic data \eqref{expabos}-\eqref{expafer} they read
\begin{equation}
\delta\beta_j  =  \bar \xi \gamma_5 \dot \beta^\psi_j + \dot{\bar \xi}  \gamma_5  \beta^\psi_j, \qquad  \delta\alpha_j =   \bar \xi \alpha^\psi_j
\qquad
\delta\beta^\psi_j = \beta_j  \xi, \qquad
\delta\alpha^\psi_j  =   \dot\alpha_j \gamma_5 \xi,
\nonumber
\end{equation}
with  $\xi(t)=\xi_0+t\xi_1$ now being  Majorana and subject to the projection $i\gamma_1 \xi = \xi$. Each of the (constant) spinors $\xi_0,\xi_1$ account  for one real supersymmetry.} multiplets $\Phi_i=(\phi_i,\psi_i)$ ($i=1,2$)
it is not difficult to see that the set of boundary conditions
 \begin{alignat}{2}
 \chi  {\alpha}_1-\beta_2 & = 0, &
 \tilde\chi  {\alpha}_2-\beta_1&= 0,
 \label{lesssusybos}
   \\
  \qquad \chi \alpha_1^\psi - \gamma_5 \dot\beta_2^\psi
  & = 0, & \qquad
 \quad  \tilde\chi \alpha_2^\psi -\gamma_5 \dot\beta_1^\psi &= 0,  \qquad 1~\text{supersymmetry}
 \label{lesssusyfer}
\end{alignat}
is invariant under  supersymmetry transformations generated by $\xi_0$, which is now Majorana. Being less supersymmetric, these boundary conditions involve two arbitrary mixing parameters $\chi$ and $\tilde \chi$. In the following sections we shall consider the case $\tilde\chi = \chi$ and refer to it as {\sf Type Ib}. The motivation for this specific choice is that the boundary conditions can be implemented by the addition of a simple boundary term to the action \eqref{action}.


\subsubsection{Mixed Boundary Conditions Involving Derivatives}
\label{sec:DerBC}


There are more interesting possibilities preserving four real supersymmetries, which is the number of supersymmetries preserved by the Wilson loops of interest to us. To discover them, notice that preserving the {\it Dirichlet} fermionic boundary condition $\alpha^\psi =0$ under supersymmetry requires actually a less stringent scalar field condition, namely $\dot\alpha=0$ (cf. \eqref{fermionbc}). Thus, we examine a mixed condition  between  $\beta = 0$ and $\dot\alpha=0$. More precisely,
\begin{equation}
 i\chi \dot\alpha - \beta = 0.
\label{derivativemixbos}
\end{equation}
Now, the supersymmetry variation of the bosonic boundary condition \eqref{derivativemixbos} vanishes provided we impose on the  fermion
\begin{equation}
 i\chi \alpha^\psi -\gamma_5 \beta^\psi = 0 .
\label{derivativemixfer}
\end{equation}
In contrast with the previous case, conditions \eqref{derivativemixbos} and \eqref{derivativemixfer} are supersymmetric without any restriction on the Killing spinor and therefore preserve as many supersymmetries as Neumann and Dirichlet boundary conditions.  Additionally, the interpolating parameter $\chi$ is dimensionless and, as will be we shown in the next section, scale symmetry in the dual 1-dimensional defect theory will be preserved.

Therefore, we have found
\begin{equation}
 \textsf{Type II:} \qquad i\chi \dot\alpha - \beta  = 0, \quad  i\chi \alpha^\psi -\gamma_5 \beta^\psi =0, \qquad 4~\text{supersymmetries}
 \label{M2}
 \end{equation}
When considering real and imaginary components of the fields  we get\footnote{Had we not introduced  the imaginary unit in \eqref{derivativemixbos}, we would have obtained $\chi \dot{\alpha}_j - \beta_j = 0$. It is not clear what boundary term could enforce such a condition, since $\chi \dot\phi \phi$ is a total time derivative.}
 \begin{alignat}{2}
\chi \dot{\alpha}_1-\beta_2  & = 0, &
  \chi \dot{\alpha}_2+\beta_1 = 0, &
  \label{Neubc2bos}
   \\
  \chi \alpha_1^\psi - \gamma_5 \beta_2^\psi & = 0, &
  \qquad
  \chi \alpha_2^\psi +\gamma_5 \beta_1^\psi  = 0. &\label{Neubc2fer}
\end{alignat}

Still, the mixed boundary condition \eqref{M2} is not an interpolation of the kind required, since condition $\dot\alpha =0$ rather than $\alpha =0$ sits at one of the endpoints. {More precisely, boundary conditions \eqref{M2} admit constant mode solutions, $\phi=const.$, for any value of $\chi$, which should be associated with delocalized configurations, and that is not what we expect for the dual description of the interpolating Wilson loop family.} In the next example we consider adding an additional constraint to the {\sf Type II}, {so that  the limiting case} of the interpolation reduces to the standard Dirichlet condition $\alpha = 0$.


\subsubsection{Mixed Boundary Conditions Without Constant Modes}
\label{sec:intBC}


{In what follows, we complement \eqref{M2} so that  constant modes are removed. One way of achieving this is to impose, as an additional constraint, that functions $\alpha_i$ vanish at some specific point. For example, we can additionally demand that}
\be
\alpha_1(t_0) = 0,\qquad \alpha_2(\tilde t_0) = 0.
\label{eq:additional}
\ee

Proceeding as before, it is straightforward to check that  conditions \eqref{eq:additional} in combination with \eqref{M2} are supersymmetric,  if we further impose that
\be
\beta_2(t_0) = 0,\quad \beta^\psi_2(t_0) = 0,\qquad
\beta_1(\tilde t_0) = 0,\quad \beta^\psi_1(\tilde t_0) = 0.
\label{eq:additional2}
\ee

Thus, our last example of supersymmetric boundary conditions is
 \begin{alignat}{3}
 &\textsf{Type III:}&  \qquad \chi \dot{\alpha}_1(t)-\beta_2(t)  & = 0, &
  \chi \dot{\alpha}_2(t)+\beta_1(t) = 0, &
  \qquad 4~\text{supersymmetries}
  \label{M3a}
   \\
&&
\alpha_1(t_0) &= 0,&
\alpha_2(\tilde t_0) = 0,&
\label{M3b}
   \\
&&
\beta_2(t_0) &= 0,&
\beta_1(\tilde t_0) = 0,&
\label{M3c}
\\
&&  \chi \alpha_1^\psi(t) - \gamma_5 \beta_2^\psi(t) & = 0, &
  \qquad
  \chi \alpha_2^\psi(t) +\gamma_5 \beta_1^\psi(t)  = 0. &
  \label{M3d}
     \\
&&
\beta_2^\psi(t_0) &= 0,&
\beta_1^\psi(\tilde t_0) = 0,&
\label{M3e  }
\end{alignat}

\textsf{Type III} boundary conditions   meet all the requirements {so far}: they are supersymmetric for unrestricted Killing spinor transformations, the interpolating parameter is dimensionless and the endpoints are precisely the Dirichlet and Neumann boundary conditions. {In the following sections, motivated by the $t$-translation invariance of the  Wilson loops we would like to describe,  we will consider the specific case of $t_0$ and $\tilde t_0$ going to $\pm\infty$. Furthermore, it is for this case that we will be able to identify an appropriate boundary term that added to the action implements the additional constraints \eqref{M3b}.} {With this necessary ingredients, we will compute 2-point correlators holographically and confirm our expectation that  boundary conditions \eqref{M3a}-\eqref{M3c} do not break the conformal invariance in the 1-dimensional theory. All this}  will lead us to propose that the supersymmetric boundary conditions of {\sf Type III} give the holographic description of the family of Wilson loops interpolating between the bosonic 1/6 BPS loop and 1/2 BPS one.


\subsection{Boundary Terms and Well Posed Variational Problems}


The different  boundary conditions discussed so far should arise from the vanishing of boundary terms in a well posed variational problem. To achieve this, the action \eqref{action} must be supplemented with appropriate boundary terms. The specification of these boundary terms is important when it comes to the evaluation of the on-shell action in order to compute correlation functions in the dual 1-dimensional defect theory (see \cite{Minces:1999eg} for related work).


\subsubsection{Type I: Standard Mixed Boundary Conditions}


To implement the first example, {\sf Type I} boundary conditions, we add the following boundary term to the action
\eqref{action}
\bea
 S_{\sf bdry}^{{\sf I}}=\frac{\chi_b}{2}\int\limits^{\infty}_{-\infty}dt\sqrt{\g}\left.\f^* \f\right|_{y=\e},
 \label{BosBdy}
\eea
with $\gamma$ the induced metric on the $y=\epsilon$ surface. The bosonic piece in the variation of the total action, after imposing the equations of motion, reads
\bea
\del (S_{\sf bulk}^ B+S_{\sf bdry}^{{\sf I}} )=\frac12
\int\limits_{-\infty}^{\infty} dt \sqrt{\g}\,
\Big[ \del\f^*(\pa_n\f+\chi_b\f) +
 (\pa_n\f^*+\chi_b\f^*)\del\f
\Big]_{y=\e}.
\label{bDy}
\eea
Here $\partial_n\phi=n\cdot\partial\phi$ stands for the (outer) unit normal derivative to the boundary. To make \eqref{bDy} vanish we  impose
\bea
\left.(\chi_b \phi+\pa_n\phi)\right|_{y=\e}=0,
\eea
which is exactly the boundary condition \eqref{standardmix} upon inserting  the expansion \eqref{expabos}\footnote{The relation between mixing parameters is
$\chi_b =\epsilon \chi+ {\cal O}(\epsilon^2)$, see \cite{Hartman:2006dy}.
}.

We now turn to the fermionic piece. Following \cite{Henneaux:1998ch}, the variation of the fermionic action evaluated on-shell takes the form
\bea
 \del S_{\sf bulk}^{F}&=&\frac i2 \int\limits^{\infty}_{-\infty}dt \sqrt{\g}\, \bar{\psi}\gamma^{n}\del\psi
 = \frac12\int\limits^{\infty}_{-\infty}dt
 ( \bar{\beta}^{\psi}\del \alpha^{\psi}-\bar{\alpha}^{\psi}\del \beta^{\psi}),
 \label{delF}
 \eea
here $\gamma^n=n\cdot\gamma$. This variation vanishes when fermions satisfy \eqref{FbcS}, indeed
\bea
 \del S_{\sf bulk}^{F}&=&  \frac1{2\chi}\int\limits^{\infty}_{-\infty}dt
 ( \bar{\beta}^{\psi}\gamma_5\del \dot\beta ^{\psi}+\dot{\bar\beta}^{\psi}\gamma_5\del \beta^{\psi})= \frac1{2\chi}\int\limits^{\infty}_{-\infty}dt\frac d{dt}\left({\bar\beta}^{\psi}\gamma_5\del \beta^{\psi}\right)=0
 \eea
Hence, the action \eqref{action} supplemented by the boundary term \eqref{BosBdy} is the appropriate action for {\sf Type I} boundary conditions. No fermionic boundary term is needed.

We conclude this subsection, briefly discussing the less supersymmetric example {\sf Type Ib} at the end of section \ref{sec:StanBC}. Consider a pair of real scalar fields satisfying
\bea
\label{eq:RobinBC1}
\chi_b\f_1+\pa_n\phi_2=0,\qquad  \chi_b\f_2+\pa_n\phi_1=0,
\eea
which is condition \eqref{lesssusybos} for $\tilde\chi = \chi$. For these, the appropriate boundary term is
\bea
 {S}_{\sf bdry}^{{\sf Ib}}
 ={\chi_b}\int\limits^{\infty}_{-\infty}dt\sqrt{\g}\,\f_1\f_2.
 \label{btlesssusy}
\eea


 \subsubsection{Type II: Mixed Boundary Conditions Involving Derivatives}


{\sf Type II} boundary conditions are obtained by adding the following boundary term to the action \eqref{action},
 \bea
 S^{{\sf II}}_{\sf{bdry}}=\frac {i\chi_b}2\int\limits_{-\infty}^{\infty}\left. d t\sqrt{\g}\,\phi^*\pa_t\phi\right|_{y=\e}.
 \label{bB}
 \eea
Indeed, after imposing the equations of motion, the bosonic piece one obtains is
 \bea
 \del (S_{\sf bulk}^B+ S^{\sf II}_{\sf bdry})=\frac12\int\limits_{-\infty}^{\infty} dt \sqrt{\g}\big(  \del \phi^*(\pa_n\phi+i \chi_b\pa_t\phi)+\del \phi(\pa_n\phi^*-i \chi_b\pa_t\phi^*)\big).
\label{Sat}
 \eea
Thus, the variational problem   becomes well defined if we demand
\be
\left.\left(\pa_n\phi+i \chi_b\pa_t\phi\right)\right|_{y=\e}=0.
\label{Bder}
\ee
This equation reproduces \eqref{derivativemixbos} when we take into account that $\chi_b=\epsilon\chi+{\cal O}(\epsilon^2)$ and   \eqref{expabos}. Again,  no boundary term is required for the fermion sector since fields satisfying \eqref{derivativemixfer} make  \eqref{delF} identically zero. Summarizying, the action \eqref{action} supplemented by the boundary term \eqref{bB} is the appropriate action for {\sf Type II} boundary conditions.


 \subsubsection{Type III: Mixed Boundary Conditions Without Constant Modes}


{As defined above, {\sf Type III} boundary conditions amount to impose some additional constraints to the the {\sf Type II} boundary conditions. We are interested in the case in which \eqref{Bder} is supplemented with  $\phi_1(-\infty,\e) = \phi_2(+\infty,\e) = 0$. These can be conveniently presented as}
\bea
\chi \phi_1+\int\limits_{-\infty}^{t}dt'\sqrt{\g}\,\pa_n\f_2 \Big|_{y=\e}=0,\quad\text{and}\quad\chi \phi_2+\int\limits_{t}^{\infty}dt'\sqrt{\g}\,\pa_n\f_1\Big|_{y=\e}=0,
\label{eq:realscalbcint}
\eea
{which can be derived from the following boundary term,}
\bea
S_{\sf{bdry}}^{\sf III}=\frac{1}{\chi}\int\limits_{-\infty}^{\infty} dt \sqrt{\g} \int\limits_{-\infty}^{t}dt'\sqrt{\g}\,
\left.\pa_n\f_1(t,y)\pa_n\f_2(t',y)  \right|_{y=\e}.
\eea
Indeed,
\begin{align}
\label{eq:varintBC}
\del(S_{\sf bulk}^B+S_{\sf{bdry}}^{\sf III})=&\int\limits_{-\infty}^{\infty} dt \sqrt{\g}\,\left(\del\f_1\,\pa_n\f_1+\del\f_2\,\pa_n\f_2\right)\\
& +\frac{1}{\chi}\int\limits_{-\infty}^{\infty} \!\! dt \sqrt{\g}
\int\limits_{-\infty}^{t}dt'\sqrt{\g}
\left[\delta\big( \pa_n\f_1(t,y)\big)\pa_n\f_2(t',y) + \pa_n\f_1(t,y)\delta\big(\pa_n\f_2(t',y)\big)\right]_{y=\e}\nonumber\\
=&\!\int\limits_{-\infty}^{\infty} \!\! dt \sqrt{\g} \left[\pa_n\f_1\, \del\!\Big(\f_1+\frac{1}{\chi}\!\int\limits_{-\infty}^{t}\!dt' \sqrt{\g}\,\pa_n\f_2 \Big)  +\pa_n\f_2 \,\del\!\Big(\f_2 +\frac{1}{\chi}\!\int\limits^{\infty}_{t}\!dt'\sqrt{\g}\,\pa_n\f_1  \Big)\right]_{y=\epsilon}.\nonumber
\end{align}
Hence, imposing \eqref{eq:realscalbcint} makes the variations in the third line to vanish\footnote{In passing to the third line we used  the identity $\int\limits^{\infty}_{-\infty} dt \,F_1(t)\int\limits^{t}_{-\infty} dt' F_2(t')=\int\limits^{\infty}_{-\infty}dt \,F_2(t)\int\limits_{t}^{\infty} dt' F_1(t') $.}. For the fermion sector no boundary terms are required,   the discussion in the previous subsection applies to the present  case.


\section{Correlators}
\label{correlators}

In this section we compute holographically, for the different types of boundary conditions discussed above, 2-point correlation functions in the dual 1-dimensional defect. We will obtain correlation functions for bosonic operators performing the canonical GKPW procedure in Euclidean space (see \cite{Gubser:1998bc},\cite{Witten:1998qj}). We will be mainly interested in studying whether scale invariance is broken or not. More concretely, we will solve the Klein-Gordon equation for scalar fields in $AdS_{2}$ as a function of arbitrary sources $f_i(\tau)$ located at the boundary.
The on-shell action, as a functional of sources, gives the generating function for $n$-point correlators in the dual QFT$_1$ in the strong coupling limit.   Two point correlators will be obtained by computing functional derivatives of it. Our work can be understood as an extension of the analysis presented in \cite{Hartman:2006dy} to  a pair of real massive scalars and to the different boundary conditions we have described in previous sections.

We work in Euclidean $AdS_{2}$,  written in Poincar\'e coordinates as in \eqref{eq:PoinPatch},  so many expressions are going to be re-written replacing $t$ by $\tau$. We regularize the problem in the standard way, {\it i.e.} by setting  the boundary conditions at $y=\e>0$ and taking the $\epsilon\to0$ at the end of the computations.  From now on, only the bosonic part of the action will be relevant, decomposing the complex scalar field into its real and imaginary parts $\phi=\phi_1+i\phi_2$ we have
\bea
S^B=\frac{1}{2}\int  d^2x\sqrt{h}  \left(h^{\a \b}(\pa_\a\phi_1 \pa_\b \phi_1+ \pa_\a\phi_2 \pa_\b \phi_2)
+m_B^2 (\phi_1^2 +\phi^2_2)\right)
+ S_{\sf bdry}.
\eea
From this action one gets the usual Klein-Gordon equations of motion for scalar fields
\bea
(\Box -m_B^2)\phi_i=0.
\label{KGeoms}
\eea
These have to be solved, in a well posed variational problem, with boundary conditions derived from enforcing
$\delta S_{\sf bulk}+\delta S_{\sf bdry}=0$
in each of the cases, as discussed in the previous section.

A Fourier transform in the temporal variable,
\bea
\phi_{i}(\t,y)=\frac{1}{\sqrt{2\pi}}\int\limits_{-\infty}^{\infty} d\w\,\tilde\phi_{i}(\w,y) e^{i \w \t}.
\eea
 translates equation \eqref{KGeoms} into
\bea
\left[y^2 \frac{d^2}{dy^2} - \w^2 y^2 - m_B^2\right]\tilde\phi_{i}(\w,y)=0.
\eea
Its  solution is well known to be given in terms of a Bessel functions \cite{Gubser:1998bc,Freedman:1998tz}
\bea
\tilde\phi_i(\w,y) = y^{1/2} K_\nu(|\w|y) A_i(\w),\qquad \nu = \sqrt{\tfrac{1}{4}+m_B^2}\,.
\label{generalsol}
\eea
Here $A_i$ are arbitrary coefficients to be fixed by the boundary source $f_i$. The case of interest to us is particularly simple, since for massless scalar fields $\nu = 1/2$ and the Bessel function reduces to an exponential.

\subsection{Type I: Standard Mixed Boundary Conditions}
The first example we discuss is that of two real scalars satisfying  {\sf Type Ib} boundary condition, now with arbitrary sources
\bea
\label{eq:RobinBCsour}
\chi_b\phi_1+ \pa_n\phi_2\big|_{y=\e}={ \epsilon}f_{1}(\t)\qquad \text{and}
\qquad  \chi_b\phi_2 + \partial_n\phi_1\big|_{y=\e}= { \epsilon} f_{2}(\t).
\eea
The corresponding boundary term, modified to account for the sources, reads
\bea
S^{{\sf Ib}}_{\sf bdry}  = \int\limits_{-\infty}^{\infty} d\t \sqrt{\g}\left.\left(\chi_b \phi_1\phi_2 - \epsilon\phi_1 f_2(\tau) -  \epsilon \phi_2 f_1(\tau)\right)\right|_{y=\e},
\eea
The boundary conditions \eqref{eq:RobinBCsour} fix the coefficients $A_i$ in \eqref{generalsol} to be
\bea
A_1(\w)= \frac{-|\w| \tilde f_{2}(\w)+\chi \tilde f_{1}(\w)}{\chi^2- \w^2} {\sqrt{\frac{2|\omega|}{\pi}}} e^{|\w|\e},
\qquad
A_2(\w)= \frac{-|\w|\tilde f_{1}(\w)+\chi \tilde f_{2}(\w)}{\chi^2- \w^2}{\sqrt{\frac{2|\omega|}{\pi}}} e^{|\w|\e},
\eea
where $\tilde f_i(\omega)$ are the Fourier transforms of the sources.

The total on-shell action, as a functional of the sources, becomes
\be
S^{B,{\sf Ib}}[f_{1},f_{2}] =\frac{1}{2}\int\limits_{-\infty}^{\infty} d\w\,
\frac{ {|\w|\tilde f_{1}(\w)\tilde f_{1}(-\w)}+{|\w|\tilde f_{2}(\w)\tilde f_{2}(-\w)}
-2\chi \tilde f_{1}(\w)\tilde f_{2}(-\w) }{\chi^2-\w^2} .
\ee
Thus, for the Fourier transform of the of the 2-point correlation functions we get
\bea
\label{noncon1}
\langle\tilde\co_1(\w_1)\tilde\co_1(\w_2) \rangle = \langle\tilde\co_2(\w_1)\tilde\co_2(\w_2) \rangle
&\!=\!&
-\frac{\del(\w_1+\w_2)|\w_1|}{\chi^2-\w_1^2},
\\
\langle\tilde\co_1(\w_1)\tilde\co_2(\w_2) \rangle=\langle\tilde\co_2(\w_1)\tilde\co_1(\w_2) \rangle
&\!=\!&
\frac{\chi\del(\w_1+\w_2)}{(\chi^2-\w_1^2)}.
\label{noncon2}
\eea
As one might have expected, due to the introduction of the dimensionful parameter $\chi$, the correlators are not scale invariant unless one considers them in the $\chi\to0$ or the $\chi\to\infty$ limits. Indeed, one gets
\bea
\langle\co_1(\t_1)\co_1(\t_2) \rangle \simeq
\left\{
\begin{array}{ll}
-\frac{1}{\pi}(\log \left|\t_1-\t_2\right| +\gamma_E )
& \qquad{\rm for}\ \chi\to 0
\\
 \frac{1}{\pi} \frac{1}{(\t_1-\t_2)^2 \chi ^2}
&\qquad {\rm for}\ \chi\to \infty
\end{array}
\right.
\eea
Thus, as $\chi$ goes from 0 to $\infty$ one finds an interpolation between CFT$_1$ 2-point functions of scalar operators with dimension $\Delta = 0$ and $\Delta = 1$, corresponding to Neumann and  Dirichlet quantizations respectively. The logarithmic behavior in the 2-point function for $\chi\to 0$ arises from IR divergences, so the appropriate quantum operators  are not $\co_{i}$ themselves but their time derivatives \cite{Polchinski:2011im} .

\subsection{Type II: Mixed Boundary Condition Involving Derivatives}

The appropriate boundary term,  enforcing boundary conditions
\bea
\label{eq:DerivBCsour}
\chi_b\pa_\t\phi_1+ \pa_n\phi_2\big|_{y=\e}= \e f_{1}(\t),\qquad
\qquad  \chi_b \pa_\t\phi_2-\partial_n\phi_1\big|_{y=\e}= \e f_{2}(\t),
\eea
written in terms of two real scalar fields is
\bea
S_{\sf bdry}^{\sf II}  = \int\limits_{-\infty}^{\infty} d\t \sqrt{\g}\,\left.\left(\chi_b\, \phi_2\pa_\t \phi_1 + \e \phi_1 f_2(\tau) -\e \phi_2 f_1(\tau)\right)\right|_{y=\e}.
\eea
The coefficients $A_i(\omega)$ in this case result
\bea
A_1(\w) = \frac{-|\w|\tilde f_2(\w)+i \chi \w \tilde f_1(\w)}{\w^2(1-\chi^2)}
\sqrt\frac{2|\omega|}{\pi}e^{\e|\w|},
\quad
A_2(\w) = \frac{ |\w|\tilde f_1(\w)+i \chi \w \tilde f_2(\w)}{\w^2(1-\chi^2)}
\sqrt\frac{2|\omega|}{\pi}e^{\e|\w|},
\eea
and the on-shell action takes the form
\bea
S^{B,{\sf II}}[f_1,f_2]=-\frac{1}{2}\int\limits_{-\infty}^{\infty} d\w\, \frac{ |\w| \tilde f_1(\w) \tilde f_1(-\w)+ |\w|\tilde f_2(\w)\tilde f_2(-\w) -2i \chi \w \tilde f_1(\w)\tilde f_2(-\w)}{\w^2(1- \chi^2)}.
\eea
The 2-point correlation functions associated to the sources $f_1$ and $f_2$ are
\bea
\langle\tilde\co_1(\w_1)\tilde\co_1(\w_2) \rangle = \langle\tilde\co_2(\w_1)\tilde\co_2(\w_2)\rangle &\!\!=\!\!&
\frac{\del(\w_1+\w_2)}{|\w_2|(1-\chi^2)},
\\
\langle\tilde\co_1(\w_1)\tilde\co_2(\w_2) \rangle = - \langle\tilde\co_2(\w_1)\tilde\co_1(\w_2) \rangle &\!\!=\!\!&-
i  \frac{\chi \del(\w_1+\w_2)}{\w_1(1-\chi^2)}.
\eea
In configuration space they are given by
\bea
\langle\co_1(\t_1)\co_1(\t_2) \rangle =
 \langle\co_2(\t_1)\co_2(\t_2) \rangle &\!\!=\!\!&
- \frac{(\g_E+ \log|\t_1-\t_2|)}{\pi(1-\chi^2)},
\label{nomixx}
 \\
\langle\co_1(\t_1)\co_2(\t_2) \rangle = -
\langle\co_2(\t_1)\co_1(\t_2) \rangle &\!\!=\!\!&-
\frac{1}{2}\frac{\chi\, \text{sign}(\t_1-\t_2)}{(1-\chi^2)}.
\label{mixx}
 \eea
The first line indicates that  the correlators are conformal, and correspond to  operators with dimension $\Delta =0$ for all values of $\chi$. As in the previous subsection,  time derivatives of $\co_i$ should be considered as the good quantum operators. The mixed correlators \eqref{mixx} should be seen as contact terms.

\subsection{Type III: Mixed Boundary Condition Without Constant Modes}

For our last example we consider the boundary term
\bea
S_{\sf bdry}^{\sf III} = \frac{1}{\chi} \int\limits_{-\infty}^\infty d\t \sqrt{\gamma}
\int\limits_{-\infty}^\t d\t'
\sqrt{\gamma}\left.
\pa_n \phi_1(\t,y)\pa_n \phi_2(\t',y)\right|_{y=\e}.
\eea
The variation of the action in this case was computed in \eqref{eq:varintBC} and its vanishing is achieved with boundary conditions
\bea
\label{eq:CauchyProInt}
\chi \phi_1+\int\limits_{-\infty}^{\t}d\t'\sqrt{\g}\,\pa_n\f_2\big|_{y=\e}=F_1(\t),\qquad
\chi \phi_2+\int\limits^{\infty}_{\t}d\t'\sqrt{\g}\,\pa_n\f_1\big|_{y=\e}=F_2(\t),
\eea
where $F_1(\t)$ and $F_2(\t)$ are arbitrary sources.

Solving for the coefficients $A_i$ one finds\footnote{We have discarded the evaluation in $-\infty$ since we can express it as: $\lim_{\L\rightarrow\infty}\int\limits_{-\infty}^{\infty}d\w \frac{|\w|}{\w} e^{-|\w|\e}e^{i \L \w}A(\w)$, which for any well behaved function  vanishes.}
\begin{equation}
A_1(\w)=\frac{i\,\text{sign}(\w)\tilde{F}_2(\w)+\chi \tilde{F}_1(\w)}{\chi^2-1}\sqrt{\frac{2|\omega|}{\pi}} e^{\e|\w|},\quad
A_2(\w)=\frac{-i\,\text{sign}(\w)\tilde{F}_1(\w)+\chi \tilde{F}_2(\w) }{\chi^2-1}\sqrt{\frac{2|\omega|}{\pi}} e^{\e|\w|}.
\end{equation}
Replacing in the on-shell action we get,
\bea
S^{B,\sf III}[F_1,F_2]&\!\!=\!\!&\frac{1}{2\chi}\int\limits_{-\infty}^{\infty} d\w \frac{|\w| \chi \big(
 \tilde{F}_1(\w)\tilde{F}_1(-\w)
+ \tilde{F}_2(\w)\tilde{F}_2(-\w)\big)
-2i\w\tilde{F}_1(\w)\tilde{F}_2(-\w) }{\chi^2-1}.
\eea
So, for the correlators we have
\bea
\langle\tilde\co_1(\w_1)\tilde\co_1(\w_2) \rangle=\langle\tilde\co_2(\w_1)\tilde\co_2(\w_2) \rangle &\!\!=\!\!& - \frac{|\w_1|\del(\w_1+\w_2)}{\chi^2-1},\\
\langle\tilde\co_1(\w_1)\tilde\co_2(\w_2) \rangle = -\langle\tilde\co_2(\w_1)\tilde\co_1(\w_2) \rangle
&\!\!=\!\!&
\frac{i \w_1\del(\w_1+\w_2)}{\chi(\chi^2-1)},
\eea
which in configuration space read
\bea
\langle\co_1(\t_1)\co_1(\t_2) \rangle=\langle\co_2(\t_1)\co_2(\t_2)\rangle&\!\!=\!\!& \frac{1}{\pi} \frac{1}{\chi^2-1}\frac{1}{|\t_1-\t_2|^2},
\label{nmx}\\
\langle\co_2(\t_1)\co_1(\t_2)\rangle=
-\langle\co_1(\t_1)\co_2(\t_2)\rangle &\!\!=\!\!& -\frac{\del'(\t_1-\t_2)}{\chi(\chi^2-1)}.
\label{mx}
\eea
As expected we find conformal correlators in the first line and contact terms in the second line. {It is instructive to spell the relation between correlation functions \eqref{nomixx}-\eqref{mixx} and \eqref{nmx}-\eqref{mx} which, up to contact terms, can be schematically expressed as
$$\langle\co^{\sf III}(\t_1 )\co^{\sf III}(\t_2) \rangle=\pa_{\tau_1}\pa_{\tau_2}\langle\co^{\sf II}(\t_1 )\co^{\sf II}(\t_2) \rangle.$$
This is an immediate consequence of the following relations between the corresponding on-shell actions,
$$S^{B,{\sf II}}[\dot F_1,\dot F_2]=S^{B,{\sf III}}[F_1,F_2]+\frac{\epsilon}{\chi}\int\limits_{-\infty}^\infty d\tau\sqrt\gamma\,F_1\dot F_2.$$
}

\section{Vacuum Energy}
\label{vacuum}

In this section we compute the 1-loop correction to the vacuum energy for the $AdS_2$ supermultiplets with the supersymmetric mixed boundary  conditions discussed so far. The interest in the  vacuum  energy for the fluctuation modes is due to its relation to the 1-loop partition function for open strings.  This  has  been  discussed in  several works, see for example \cite{Drukker:2000ep,Sakai:1984nc,Camporesi:1992wn}. The idea is to compare these vacuum energies for different boundary conditions. More precisely, we would like to see whether this computation depends on the interpolating parameter $\chi$ or not.

The full set of modes consists in   two scalars of mass $m^2_B=2$, six scalars of mass $m_B^2=0$, two fermions with $m_F=0$ and six other fermions with $|m_F|= 1$. Following \cite{Drukker:2000ep}, we will compute the 1-loop correction to vacuum energy by the on-shell method adding up the zero point energies of the quantum field modes in  global $AdS_2$ with metric
\be
ds^2 = \frac{1}{\cos^2\s}(d{\sf t}^2-d\s^2).
\label{globalAdS2}
\ee
The supersymmetry analysis of the different types of mixed boundary conditions we have considered so far can be naturally translated to  global coordinates. Some details are presented in appendix \ref{sec:NBexpansion}. The global time coordinate $\sf t$ appears with a different font to manifest the fact that it is not the same time coordinate as in Poincar\'e coordinates.

In \cite{Sakai:1984vm},  the set of modes for free bosons and fermions in $AdS_2$ were chosen based on fall-offs  that  guarantee the  conservation of: (i)  energy and (ii) the  bosonic and fermionic inner products. In particular, one starts with the {\it formally} conserved charge associated to the timelike Killing vector $\bm k=\partial_{\sf t}$,
\bea
 E=\int\limits_{-\frac{\pi}{2}}^{\frac{\pi}{2}} d\s\sqrt{-g} g^{\sf t\m} T_{\m\n}k^{\n} ,
\eea
where $T_{\mu\nu}$ is the energy-momentum tensor. Proper conservation is achieved by imposing the vanishing of the energy flux at spatial infinity,
\bea
 \sqrt{-g} g^{\s\m} T_{\m\n}k^{\n} \big|_{\s=\pm\frac{\pi}{2}} = 0.
\eea
For a scalar field this equation becomes equivalent to
\bea
  \label{eq:Eflux}
  -\pa_{\sf t}\f\, \pa_\s\f\big|_{\s=\pm\frac{\pi}{2}} = 0.
\eea

Demanding the scalar field to have  a single fast fall-off at the  boundary of the form $\phi\sim(\frac{\pi}{2}\pm\s)^{\Delta_+}$, loosely referred to as Dirichlet boundary condition\footnote{We denote $\Delta_{\pm} = \frac12 \pm \nu$ where $\nu = \sqrt{\frac14+m_B^2}$.}, forces the following supersymmetric spectrum of frequencies \cite{Sakai:1984vm}
\bea
 \w_B^{\,_{\sf D}}(n,m_B^2)=n+\Delta_+\qquad \text{and} \qquad \w_F^{\,_{\sf D}}(n,|m_F|)=n+|m_F|+\frac{1}{2}.
\eea
The total vacuum energy contributing to the 1-loop partition function of the  {\it Dirichlet} string, dual to the 1/2 BPS Wilson line, was computed in \cite{Aguilera-Damia:2014bqa} as
\bea
E_{1/2}^{\sf 1-loop}=\frac{1}{2}\left(\sum_{n=0}^{\infty} (2 \w_B^{\,_{\sf D}}(n,2)+6 \w_B^{\,_{\sf D}}(n,0)-2\w_F^{\,_{\sf D}}(n,0)-6\w_F^{\,_{\sf D}}(n,1)\right),
\label{TVE}
\eea
which, as usual, requires an appropriate regularization. This vacuum energy can be re-expressed in terms of the Hurwitz  $\zeta$-function,
 \bea
 \z(s,\Delta)=\sum_{n=0}^\infty (n+\Delta)^{-s}\qquad \text{and}\qquad \z(-1,\Delta)=-\frac{1}{2}\left(\Delta^2-\Delta+\frac{1}{6}\right).
 \eea
Thus, for Dirichlet boundary conditions the 1-loop correction \eqref{TVE} results
 \bea
E_{1/2}^{\sf 1-loop} &\!\!=\!\!&
\frac12\left[2\z(-1,2)+6\z(-1,1)-2\z(-1,\tfrac{1}{2})-6\z(-1,\tfrac{3}{2})\right]\nn\\
&\!\!=\!\!&
-\frac{1}{4}\left[2\times(2+\tfrac{1}{6})+6\times(\tfrac{1}{6})-2\times(-\tfrac{1}{12})-6\times(1-\tfrac{1}{12}) \right]=0.
\eea

When we turn to the 1-loop partition function of the {\it Neumann} string, dual to the bosonic 1/6 BPS Wilson line, two of the massless scalar modes have the slower fall-off $(\frac{\pi}{2}\pm\s)^{\Delta_-}$. The bosonic frequencies in the Neumann case result
\be
\w_B^{\,_{\sf N}}(n,m_B^2)=|n+\Delta_-|.
\ee
Their supersymmetric fermion partners, with $-\frac{1}{2}<m_F<\frac{1}{2}$, have frequencies
\be
\w_F^{\,_{\sf N}}(n,m_F)= n+m_F+\frac{1}{2},
\ee
For the case of interest, the modes with the Neumann boundary conditions have $m_B = m_F = 0$, thus their contribution to the total vacuum energy does not change. This implies that \cite{Aguilera-Damia:2014bqa} $$E_{1/6}^{\sf 1-loop}  = 0.$$
One finds  that  1-loop determinants for the string fluctuations dual to either the  $ {1}/{2}$  or  $ {1}/{6}$ BPS Wilson lines are trivial. The result is consistent with the fact that these Wilson lines have unit expectation values.

\subsection{Mixed Boundary Conditions }

We are now interested in computing the vacuum energies for {\sf Type II} and {\sf Type III} boundary conditions, which preserve as many supersymmetries as the Neumann  case.
We start analyzing {\sf Type II} boundary conditions \eqref{Bder}. In global coordinates they read
\bea
\label{eq:MixDerBC}
i\chi \pa_{\sf t}\phi - \pa_\s\phi \big|_{\s=\pm\frac\pi2} = 0.
\eea
The solution for a massless complex scalar in global $AdS_2$ is simply given by
\bea
\label{eq:ScCompSol}
\f({\sf t},\s)=e^{i \w\sf t} \big(A\cos(\w \s)+B \sin(\w \s) \big),
\eea
with $\w\ge0$. To find the spectrum of modes, we now proceed to impose \eqref{eq:MixDerBC} at both ends. From  the behavior at $\s=\frac\pi2$ we get a relation between integration constants,
\bea
\label{eq:Cond_r+scal}
B= \frac{ \sin \left(\frac{\pi  \omega }{2}\right)- \chi  \cos \left(\frac{\pi  \omega }{2}\right)}{\chi  \sin \left(\frac{\pi  \omega }{2}\right)+\cos \left(\frac{\pi  \omega }{2}\right)}A.
 \eea
Then,
 \bea
 \f({\sf t},\s)= A  e^{i  \omega\sf t } \frac{ \left(\chi  \sin \left(\tfrac{\w\pi}{2} -\omega\s \right)+\cos \left(\tfrac{\w\pi}{2} -\omega\s \right)\right)}{\chi  \sin \left(\frac{\pi  \omega }{2}\right)+\cos \left(\frac{\pi  \omega }{2}\right)}.
 \eea
Imposing now the boundary condition at $\s=-\frac\pi2$, we obtain,
 \bea
\frac{A \left(\chi ^2+1\right) \omega  \sin (\pi  \omega )}{\chi  \sin \left(\frac{\pi  \omega }{2}\right)+\cos \left(\frac{\pi  \omega }{2}\right)}=0.
 \eea
This implies  for {\sf Type II},
\be
\w_B^{\,_{\sf II}}(n,0)=n ~~~~\text{with}~~n=0,1,2...
\label{globB}
\ee

To analyze the fermionic sector we adopt the representation \eqref{eq:VrepGam} for  $\g$-matrices, then upper and lower components of the spinor corresponds to eigenvectors of $P_\pm$.  Solutions to the Dirac equation \eqref{dirac} are of the form
\bea
&\j({\sf t},\s)=e^{i \w\sf t}\begin{pmatrix}
\j_+(\s)\\
\j_-(\s)
\end{pmatrix},
\eea
where
\bea
\!\!
\j_+(\s)= \sqrt{\cos\s}\left(\m_1\cos(\w\s) +\m_2\sin(\w\s)\right),
\qquad
\j_-(\s)= \sqrt{\cos\s}\left(\m_2\cos(\w\s) -\m_1\sin(\w\s)\right),
\eea
and $\w>0$. The supersymmetric boundary conditions for fermions associated to \eqref{eq:MixDerBC}
are (see \eqref{eq:ferBCboth} in appendix \ref{sec:NBexpansion})
\be
\left(\chi \j_\pm \pm \j_\mp\right) \big|_{\s = \pm\frac\pi2} = 0.
\ee
Imposing the right boundary condition we get the relation,
\bea
\mu_2=
\frac{\sin\left(\tfrac{\pi\omega }{2}\right)-\chi \cos\left(\tfrac{\pi\omega }{2}\right)}{\chi \sin \left(\tfrac{\pi\omega }{2}\right)+\cos\left(\tfrac{\pi\omega }{2}\right)}\m_1.
\eea
From the left boundary we obtain,
\bea
\frac{\mu_1(1+\chi^2)\cos(\pi\omega)}
{\chi \sin \left(\tfrac{\pi\omega}{2}\right)+\cos\left(\tfrac{\pi\omega }{2}\right)}=0.
\eea
This implies
\be
\w_F^{\,_{\sf II }}(n,0)=n+\frac{1}{2}~~~~\text{with}~~n=0,1,2...
\label{globF}
\ee
From \eqref{globB} and \eqref{globF} we find that  for {\sf Type II}  boundary conditions, the frequency spectrum  for fermions and bosons becomes independent of $\chi$ and coincides with that for Neumann or Dirichlet. We thus conclude that the vacuum energy and its associated 1-loop correction are independent of the interpolating parameter $\chi$.
The  analysis we have carried is straightforwardly extended to the integrated {\sf Type III}  boundary conditions leading to the same conclusion.

Had we have considered mixed boundary conditions {\sf Type Ib} of the form
\bea
\label{eq:MixStaBC}
\chi \phi \mp \pa_\s\phi \big|_{\s=\pm\frac\pi2} = 0
\eea
we would have obtained the following condition on the scalar field frequencies
\be
\chi + \w \tan(\tfrac{\w\pi}{2}) =0\quad\text{or}\quad \chi - \w \cot(\tfrac{\w\pi}{2}) =0.
\ee
In this case, as one might have expected, the frequency spectrum  depends non-trivially on the deformation parameter.

\section{Discussion}
\label{discu}

We will now discuss several properties of the mixed boundary conditions for scalar and fermionic fields in $AdS_2$ studied in section \ref{22}. We have explored different possibilities for boundary conditions interpolating between Dirichlet and Neumann. Our aim was to identify  boundary conditions that could correspond to a family of supersymmetric Wilson loops preserving four real supercharges \cite{Ouyang:2015iza,Ouyang:2015bmy}, therefore we restricted our analysis to boundary conditions invariant under supersymmetry transformations \eqref{scalarbc}-\eqref{fermionbc}.
In particular, keeping in mind that we wanted to describe the fluctuations on an $AdS_2$ open string  world-sheet in the $AdS_4\times \mathbb{CP}^3$ background, we concentrated on massless scalar and fermionic fields. As we have seen, the existence of massless fermionic fields turned out to be crucial in order to allow for supersymmetric Neumann and mixed boundary conditions. This should be correlated to the fact that in ABJM models one finds a richer variety of supersymmetric Wilson loops, as compared to the ${\cal N}=4$ SYM case.

The first example analyzed, which we named {\sf Type I}, was supersymmetric only when restricting to transformations generated by $\xi_0$. Thus, this type of boundary conditions preserved 2 real supersymmetries. Since the 1/6 BPS  family of Wilson loops is invariant under 4 real supersymmetries, {\sf Type I} conditions of the form  $\chi \alpha -\beta =0$ were ruled out as their holographic description. Another sign of inadequacy  followed from the fact that the interpolating parameter $\chi$ is dimensionful. Explicitly, the breaking of conformal invariance in the dual field theory was revealed in the computation of holographic correlators \eqref{noncon1}-\eqref{noncon2}.

The boundary conditions termed {\sf Type II} and {\sf Type III} appeared to be much more appealing. Firstly, they preserved 4 real supersymmetries, which are as many as those in the Neumann case. Secondly, the interpolating parameter is dimensionless suggesting that  in the dual theory conformal invariance will not be broken. This was further confirmed in section \ref{correlators} by explicitly computing holographic 2-point correlation functions. Moreover, in section \ref{vacuum}, we computed the 1-loop correction to the partition function for both {\sf Type II} or {\sf Type III} boundary conditions  finding a $\chi$-independent result. These results then agree with the expectations for the Wilson loop family vev which is also independent of the interpolating parameter $\zeta$\footnote{A fact following from all the $\zeta$-deformed Wilson loops being cohomologically equivalent (see app.\ref{susyWL}).}.

Still, {\sf Type II} and {\sf Type III} are not equally good candidates. In the $\chi\to\infty$ limit of the interpolation, the {\sf Type II} case becomes   $\dot\alpha =0$ rather than $\alpha =0$ as expected for the Dirichlet endpoint. Therefore, we proposed that the holographic description of the ABJM family of interpolating Wilson loops \cite{Ouyang:2015iza,Ouyang:2015bmy} is in terms of boundary conditions for the world-sheet fluctuations of the {\sf Type III}, given in section \ref{sec:intBC}.
This is the main result of the paper.

Our proposal gives support for Neumann boundary conditions as the correct dual interpretation of the bosonic 1/6 Wilson loop. The latter preserve a  $SU(2)\times SU(2)$ rather than $SU(3)$ of the $R$-symmetry group, and it was known since its original construction that it could not correspond to a string localized at a point in $\mathbb{CP}^3$. The proposal in \cite{Drukker:2008zx} was  to {\it smear} the dual string  along a $\mathbb{CP}^1\subset\mathbb{CP}^3$, although its meaning as a boundary condition was not clear. This was later interpreted as Neumann boundary conditions along that $\mathbb{CP}^1$ \cite{Lewkowycz:2013laa}.  The results of the present paper further confirm this interpretation: Neumann boundary conditions for the $\mathbb{CP}^1$ modes preserve 4 real supersymmetries and lead to a vanishing 1-loop correction to the partition function. The 1/6 BPS boundary conditions then  interpolate between Neumann and Dirichlet as $\chi$ goes from zero to infinity.

The relation between the Wilson loop parameter $\zeta$ and the boundary condition parameter $\chi$ is still missing in our proposal. In principle, $\chi$ could be a non-trivial function of $\zeta$ and the 't Hooft coupling $\lambda$ (cf. \cite{Hartman:2006dy} for related work). Thus, in order to determine the relation, one would need a  field theory computation that depends on $\zeta$ and that could be extrapolated to the strong coupling limit.

The quest for the appropriate boundary  conditions for the dual description of certain supersymmetric interpolating Wilson loops, led us to explore different possibilities. While {\sf Type I} and {\sf Type II} might be regarded as false attempts for the original motivation, they seem to be perfectly consistent boundary conditions from a world-sheet perspective. This prompts the opposite question of whether it is possible to realize them in terms of Wilson loops. If a dual Wilson loop corresponded to a  {\sf Type I} condition, it should be a less supersymmetric one, breaking the  conformal invariance on the $d=1$ defect. We do not have a concrete proposal for it, but it would be interesting to further explore this possibility.

{\sf Type II} condition, which interpolates between $\dot\alpha=0$ and $\beta=0$, leaves the boundary value of the scalar fields delocalized for any $\chi$. Thus, a dual Wilson loop should not only be 1/6 BPS but also preserve a $SU(2)\times SU(2)$ $R$-symmetry. This might be realized  averaging the  Wilson loop family, defined in terms of \eqref{calM},  over the $SU(2)\subset SU(4)$ rotations that act in the internal space directions $I=1,2$. The $\zeta=1$ endpoint will preserve only 4 real supersymmetries, those common to all orientations. The $\zeta=0$ endpoint will correspond  to the usual bosonic 1/6 BPS loop, since in this case the $SU(2)$ averaging do not have any effect on it.

There are further interesting problems in connection  with this family of interpolating Wilson loops. One could study them perturbatively as in \cite{Beccaria:2017rbe} and  also to compute higher point holographic correlators  as in \cite{Beccaria:2019dws}. The study of correlators in defect CFT's defined by ABJM Wilson lines \cite{Bianchi:2017ozk,Bianchi:2018scb} could be extended to the whole family of interpolating Wilson loops.
Additionally, one could ask about integrability aspects of this interpolating Wilson loop not only  from the gravity side but also from the field theory as was done for the $\cn=4 $ SYM case in \cite{Correa:2018fgz}. Interpolating Wilson loops  have also been constructed in quiver Chern-Simons-matter theories \cite{Ouyang:2015bmy,Mauri:2017whf} and could also be described holographically along the lines of this article. More ambitiously, one would also like to understand the rich phase space structure  of Wilson loops in ABJM. While we have focused only on a 1-parameter family of Wilson loops associated with marginal deformations in the 1-dimensional defect,
there are many other possible interpolations. For example, one can consider interpolations in which the full $SU(4)$ $R$-symmetry is preserved at one of the endpoints,  meaning a Wilson loop coupled only to gauge fields or a matrix $\cal M$ proportional to the identity. Although we were mainly interested in investigating  dual description of Wilson loops, it is also natural to think in  using  these mixed boundary conditions as holographic double trace deformations in other setups were $AdS_2$ appears as holographic dual, for example supersymmetric  SYK models \cite{Fu:2016vas}.

\section*{Acknowledgments}

We would like to thank Jerem\'ias Aguilera-Damia for useful discussions on this problem. We also thank the referee for their thorough review, and for the many comments, corrections and suggestions that improved our presentation. This work was supported by PIP 0681, PIP 1109, PIP UE 084 {\it B\'usqueda de Nueva F\'isica},  UNLP X791 and UNLP X850.

\appendix


\section{$AdS_2$ Spinor  Conventions and Supersymmetry}
\label{SpinConv}

\subsection{Conventions}
\label{sec:2dConventions}

In this appendix we collect different conventions we use along the paper to describe spinors and supersymmetry in $AdS_2$.\\
{\sf . Poincar\'e coordinates:}
\be
ds^2 = \frac{dt^2-dy^2}{y^2}.
\ee

\noindent{\sf Frames}: $~~  {\bm e}^0=\frac{d t}{y},~~  {\bm e}^1=\frac{dy}{y}~~~\leadsto~~~$ {\sf Spin~connection}: $~\bm{ \omega}^{01}= \bm { e}^0$.\\
{\sf Covariant~derivatives}: $D_\mu=\partial_\mu+\frac14\omega_\mu{}^{ab}\gamma_{ab} ~\leadsto~~ D_{ t}=\partial_{  t}+\frac1{2y} \gamma_5, ~~~D_{y}=\partial_y ~$ with $~\gamma_5=\gamma_0\gamma_1$.\\
{\sf Curved gammas:} $ \gamma_\mu=e^a_\mu\gamma_a $. Then,
 $ \gamma_{  t}=\frac1y\gamma_0,~~\gamma_y=\frac1y \gamma_1~$  with $\{\gamma_a,\gamma_b\}=2\eta_{ab},~~\eta_{ab}={\rm diag}(1,-1)$\\
It is convenient to define projectors along the (unit) normal direction to the boundary $n=-y\partial_y$ as
\be
P_\pm\equiv\frac12(1\pm in^\a\gamma_\a)=\frac12(1\pm i\gamma_1)~~\leadsto ~~\psi_-=P_-\psi\quad\text{and}\quad \psi_+=P_+\psi.
\label{ppm}
\ee
{\sf Killing spinor equation:}
\bea
\label{eq:KSEA}
D_\mu  \varepsilon+ \frac{i}{2}\g_\mu\varepsilon=0.
\eea
We start solving the $y$ component from \eqref{eq:KSEA}
$$\partial_y \varepsilon=-\frac i{2y}\gamma_1\varepsilon~~\rightarrow~~\varepsilon({  t},y)=e^{-\frac i2   {\log y} \, \gamma_1 }\varepsilon(  t) $$
Inserting in the $t$-equation  one finds
$$ \dot\varepsilon({  t})=-\frac iy e^{i\log y\,\gamma_1 }P_+ \gamma_0\,\varepsilon({  t})~~\rightarrow~~
\left\{\begin{array}{ll}
\dot\varepsilon_-( t)= 0\\
\dot\varepsilon_+({  t})=- i\gamma_0\, \varepsilon_-({ t})
\end{array}\right.~~\rightarrow~~
\varepsilon({  t})=\left(1-i{   t}\gamma_0P_-\right)\epsilon$$
The Killing spinor ends up depending on a constant  spinor $\epsilon$ which we decomposed  as
 $\epsilon=\epsilon_{ +}+\epsilon_{ -}$ with $i\gamma_1\epsilon_{ \pm}=\pm\epsilon_{ \pm}$ gives  \cite{LPT}
\begin{align}
\varepsilon(t,y)
&= {y^{-1/2}}\epsilon_+ +\left(y^{1/2}+{y^{-1/2}}(-i{  t}\gamma_0)\right)\epsilon_-
\label{KSS}
\end{align}
Notice that there are two different types of supersymmetries:
(i) those generated by $\epsilon_+$ are independent of the boundary coordinate $ t$ and (ii)
 those coming from by $\epsilon_-$ depend on $  t$.

In the analysis of supersymmetric boundary conditions it is useful to notice that the Killing spinor can be written as
\be
\varepsilon(t,y)= {y^{-1/2}}\xi(t) + y^{1/2} i \gamma_0\, \dot\xi (t)~~~~\text{with}~~~\ddot\xi(t)=0,~~i\gamma_1 \xi(t)= \xi(t)~\leadsto~ \xi(t)=\xi_0+t\xi_1\,,
\ee
$\xi_{0,1}$ are easily related to $\epsilon_{\pm}$ in \eqref{KSS}.\\
{ \sf . Global coordinates:} we denote the time coordinate by $\sf t$ to stress the different  folliation in global coordinates
\be
ds^2 = \frac{d{\sf t}^2-d\s^2}{\cos^2\s},\qquad {\rm for}\qquad
-\frac{\pi}{2}\leq \s \leq \frac{\pi}{2}.
\ee
The solution to  \eqref{eq:KSEA}  in this case is \cite{Sakai:1984vm}
\be
\varepsilon({{\sf t},\sigma}) = \cos^{-1/2}\s \left(\cos\tfrac{\s}{2}-i\g_1\sin\tfrac{\s}{2}\right)\xi({\sf t}) ~~ {\rm where}~~
\xi({\sf t}) = \left(\cos\tfrac{\sf t}{2}-i\g_0\sin\tfrac{\sf t}{2}\right)\xi_0.\label{KSg}
\ee
with $\xi_0$ an arbitrary constant spinor.\\
{\sf Dirac gammas}: so far we have not attached ourselves to any particular  representation.
At some point we will use the representation,
\bea
\label{eq:VrepGam}
\g_0= \begin{pmatrix}
0& 1\\
1&0
\end{pmatrix},\quad\g_1= i\begin{pmatrix}
1& 0\\
0&-1
\end{pmatrix},\quad \g_5=\g_0\g_1, \quad C=\begin{pmatrix}
0& -i\\
i&0
\end{pmatrix}.
\eea


\subsection{Supersymmetry Analysis in Global $AdS_2$}
\label{sec:NBexpansion}


In global coordinates, the Klein-Gordon and the Dirac equations read,
\bea
\frac{1}{\cos^2\s}\left(\frac{\partial^2\phi}{\partial{\sf t}^2} -
\frac{\partial^2\phi}{\partial \s^2}\right) + m_B^2\,\phi &\!\!=\!\! & 0,
\\
 \cos\s\left( \gamma_0 \frac{\partial \psi}{\partial\sf t}
- \gamma_1 \frac{\partial \psi}{\partial \s}\right)
-\frac{1}{2} \sin\s \gamma_1 \psi -m_F\,\psi&\!\!=\!\! & 0.
\label{dirac}
\eea
Since there are two asymptotic  boundaries, we shall use indices $p$ and $m$ to distinguish  between the expansions of the fields near  $\s = +\frac\pi2$ and $\s =-\frac\pi2$ respectively.\\

\nin { \sf Right boundary}: the Killing spinor   takes the form
\bea
\label{eq:KS+}
\varepsilon ({\sf t,\sigma}) =
 \sqrt{2} (\tfrac{\pi}{2}-\s)^{-1/2}P_-\xi({\sf t}) +
 \tfrac{1}{\sqrt{2}} (\tfrac{\pi}{2}-\s)^{1/2}P_+\xi({\sf t})
 +{\cal O}\left((\tfrac{\pi}{2}-\s)^{3/2}\right).
 \eea
For massless scalars and Dirac fields, the asymptotic  expansions read
\bea
\label{eq:scal+}
\hspace{-1cm}\phi({\sf t},\sigma) &\!\!\!=\!\!\! &
\left(\alpha_p({\sf t})+\tfrac{1}{2} \ddot\alpha_p( {\sf t})(\tfrac{\pi}{2}-\s)^{2} +\cdots\right) + (\tfrac{\pi}{2}-\s)
\left(\beta_p( {\sf t}) +\tfrac{1}{6} \ddot\beta_p( {\sf t})(\tfrac{\pi}{2}-\s)^{2} +\cdots \right)
\\
\label{eq:Spi+}
\hspace{-1cm}\psi ({\sf t},\sigma) &\!\!\!=\!\!\! & (\tfrac{\pi}{2}-\s)^{\frac12}\left(\alpha^{\psi}_p({\sf t})
- (\tfrac{\pi}{2}-\s)\g_5\dot\alpha^\psi_p({\sf t})+...\right)
+(\tfrac{\pi}{2}-\s)^{\frac12}\left(\beta^{\psi}_p({\sf t})
- (\tfrac{\pi}{2}-\s) \g_5\dot\beta^\psi_p({\sf t})+...\right)
\eea
where, as in Poincar\'e coordinates, $P_-\alpha_p^\psi = \alpha_p^\psi$ and $P_+\beta_p^\psi = \beta_p^\psi$.\\
\nin { \sf Left boundary}: for the Killing spinor we have
 \bea
 \label{eq:KS-}
\varepsilon({\sf t},\sigma)=
 \sqrt{2} (\tfrac{\pi}{2}+\s)^{-1/2}P_+\xi({\sf t}) +
 \tfrac{1}{\sqrt{2}} (\tfrac{\pi}{2}+\s)^{1/2}P_-\xi({\sf t})
 +{\cal O}\left((\tfrac{\pi}{2}-\s)^{3/2}\right).
 \eea
While for massless Klein-Gordon and Dirac fields we obtain
\bea
\label{eq:scal-}
\hspace{-1cm}\phi({\sf t},\s) &\!\!\!=\!\!\! &
\Big(\alpha_m({\sf t})+\tfrac{1}{2} \ddot\alpha_m({\sf t})(\tfrac{\pi}{2}+\s)^{2} +...\Big) + (\tfrac{\pi}{2}+\s)
\left(\beta_m({\sf t}) +\tfrac{1}{6} \ddot\beta_m({\sf t})(\tfrac{\pi}{2}+\s)^{2} +... \right)
\\
\label{eq:Spi-}
\hspace{-1cm}\psi ({\sf t},\s) &\!\!\!=\!\!\! & (\tfrac{\pi}{2}+\s)^{\frac12}\left(\alpha^{\psi}_m({\sf t})
+ (\tfrac{\pi}{2}+\s)\g_5\dot\alpha^\psi_m({\sf t})+...\right)
+(\tfrac{\pi}{2}+\s)^{\frac12}\left(\beta^{\psi}_m({\sf t})
+ (\tfrac{\pi}{2}+\s) \g_5\dot\beta^\psi_m({\sf t})+...\right)
\eea
{\sf Susy transformations}: from \eqref{susytrans} we get:

 $\s\to+\frac{\pi}{2}$:
\bea
&\del \a_p = \sqrt2\bar\xi  \b^\j_p ,\quad & \quad
\del \b_p  =- \sqrt2\frac{d}{d{\sf t}}\left(\bar\xi  \g_5 \a^\j_p \right)
\label{glo1}
\\
&\del \a^{\j}_p = -i\sqrt2 \b_p P_-\xi ,
\quad & \quad
\del \b^{\j}_p = i\sqrt2 \dot{\a}_p \g_5 P_-\xi .
\eea

 $\s\to-\frac\pi2$,
\bea
&\del \a_m = \sqrt2\bar\xi  \a^\j_m ,\quad & \quad
\del \b_m  = \sqrt2\frac{d}{dt}\left(\bar\xi  \g_5 \b^\j_m \right)
\\
&\del \a^{\j}_m  = -i\sqrt2 \dot{\a}_m  \g_5 P_+\xi ,
\quad & \quad
\del \b^{\j}_m = -i\sqrt2 {\b}_m   P_+\xi ,\label{glo2}
\eea
where we have used that $\dot\xi({\sf t}) = -\frac{i}{2}\g_0 \xi({\sf t})$,  cf. \eqref{KSg}.\\

\nin {\sf Susy invariance of boundary conditions}: we impose in both boundaries\footnote{The relative orientations between time and radial derivatives have to be the same in both boundaries. This enforces the same relative sign for the terms in the boundary conditions at $\frac\pi2$ and $-\frac\pi2$ in \eqref{eq:derBCC}.}
\bea
\label{eq:derBCC}
i\chi \pa_t\phi - \pa_\s\phi \big|_{\s=\pm\frac\pi2} = 0,
\eea
which in terms of the expansion coefficients give
\be
\label{eq:compBC}
i\chi \dot\a_p  +\b_p  =0,\qquad i\chi \dot\a_m  -\b_m =0.
\ee
Acting with \eqref{glo1}-\eqref{glo2} on these equations, we find they are preserved  if
\bea
\label{eq:ferBCboth}
i\chi \b^\j_p  -\g_5 \a^\j_p  =0,
\qquad
i\chi \a^\j_m  -\g_5 \b^\j_m  =0,
\eea
with  no constraint on the Killing spinor. We conclude that boundary conditions \eqref{eq:compBC}-\eqref{eq:ferBCboth} preserve all the supersymmetries generated by $\varepsilon$.

The integrated boundary conditions in global coordinates are
\be
\label{eq:compBCint}
i\chi \a_p({\sf t}) + \int\limits_{-\infty}^{\sf t} d{\sf t}' \b_p({\sf t} ') =0,
\qquad
i\chi \a_m({\sf t}) - \int\limits_{-\infty}^{\sf t} d{\sf t}' \b_m(\sf t') =0,
\ee
which are shown to preserve  all the supersymmetries when accompanied
with fermionic boundary conditions  \eqref{eq:ferBCboth}.

\section{Supersymmetric Wilson Loops Family in ABJM}
\label{susyWL}


Supersymmetric ABJ(M) Wilson loops can be expressed in terms of a $U(N|M)$ superconnection $L$ \cite{Drukker:2009hy,Lee:2010hk,Cardinali:2012ru},
\be
W_{\cal R} = {\rm tr}_ {\cal R} {\cal P} e^{i \oint   L d\tau}
\ee
where,
\be
L =\begin{pmatrix}
A_\mu \dot{x}^\mu-\frac{2\pi i}{k}|\dot{x}|\mathcal{M}^{I}_J C_I \bar{C}^J & -i\sqrt{\frac{2\pi}{k}}|\dot x|\eta^{\alpha}_{I} \bar{\psi}^{I}_{\alpha}
\\
 -i\sqrt{\frac{2\pi}{k}}|\dot x|\bar{\eta}^{I}_{\alpha} \psi^{\alpha}_{I} & \hat{A}_\mu \dot{x}^\mu-\frac{2\pi i}{k}|\dot{x}|\hat{\mathcal{M}}^{I}_J \bar{C}^J C_I
\end{pmatrix}.
\ee
For a straight line
\be
x^\mu(\tau) = (0,0,\tau),
\ee
and taking
\be
{\cal M} = \hat{\cal M} = {\rm diag}(-1,-1+2\zeta^2,1,1),\qquad
\eta^{\alpha}_{I}=\z \eta\,  \delta^{\alpha}_{+}\delta^{1}_{I},
\qquad
\bar{\eta}^{I}_\alpha=\z \bar{\eta}\, \delta^{+}_{\alpha}\delta^{I}_{1},
\label{calM}
\ee
with $\eta  \bar\eta = 2 i$ \cite{Ouyang:2015iza,Ouyang:2015bmy}, one describes a 1-parameter family of supersymmetric Wilson loops. This family interpolates between the   bosonic 1/6 BPS Wilson loop for $\zeta =0$ and the 1/2 BPS Wilson loop  for $\zeta = 1$. For generic values of $\zeta$, the Wilson loop preserves the same supercharges as the bosonic 1/6 Wilson loop and $U(1)\times U(1)\times SU(2)\subset SU(4)$.

The expectation value of  the bosonic $1/6$ straight Wilson line is equal to 1. Thus, the fact that the difference $W^{(\z)}_{\cal{R}}-W^{(0)}_{\cal{R}} $ is  $\cq$-exact for all values of $\zeta$ \cite{Ouyang:2015bmy} implies that the whole family has trivial expectation value.

\bibliographystyle{JHEP}

\begin{thebibliography}{10}

\bibitem{Rey:1998ik}
  S.~J.~Rey and J.~T.~Yee,
  \textit{``{Macroscopic strings as heavy quarks in large N gauge theory and anti-de Sitter supergravity}''},
   \textsf{Eur.~Phys.~J.~C 22, ~379 ~ (2001)},
   \href{http://arXiv.org/abs/hep-th/9803001}{\texttt{hep-th/9803001}}.

\bibitem{Maldacena:1998im}
J.~M.~Maldacena,
\textit{``{Wilson loops in large N field theories}''},
\textsf{Phys.~Rev.~Lett.~80,~4859~(1998)},
\href{http://arXiv.org/abs/hep-th/9803002}{\texttt{hep-th/9803002}}.

\bibitem{Drukker:1999zq}
  N.~Drukker, D.~J.~Gross and H.~Ooguri,
  \textit{``Wilson loops and minimal surfaces''},
  \textsf{Phys.\ Rev.\ D {\bf 60}, 125006 (1999)},
 \href{http://arXiv.org/abs/hep-th/9904191}{\texttt{hep-th/9904191}}.


\bibitem{Alday:2007he}
  L.~F.~Alday and J.~Maldacena,
  \textit{``Comments on gluon scattering amplitudes via AdS/CFT''},
  \textsf{JHEP~{\bf 0711},~068~(2007)},
  \href{http://arXiv.org/abs/0710.1060}{\tt arXiv:0710.1060}

\bibitem{BF}
  P.~Breitenlohner and D.~Z.~Freedman,
  \textit{Positive Energy in anti-De Sitter Backgrounds and Gauged Extended Supergravity''},
  \textsf{Phys.\ Lett.\  {\bf 115B}, 197 (1982)};
  P.~Breitenlohner and D.~Z.~Freedman,
  \textit{``Stability in Gauged Extended Supergravity''},
  \textsf{Annals Phys.\  {\bf 144}, 249 (1982)}.

\bibitem{Witten:2001ua}
  E.~Witten,
  \textit{``Multitrace operators, boundary conditions, and AdS/CFT correspondence''},
  \href{http://arXiv.org/abs/hep-th/0112258}{\texttt{hep-th/0112258}}.

\bibitem{Klebanov:1999tb}
  I.~R.~Klebanov and E.~Witten,
  \textit{``AdS / CFT correspondence and symmetry breaking''},
  \textsf{Nucl.\ Phys.\ B {\bf 556} (1999) 89},
  \href{http://arXiv.org/abs/hep-th/9905104}{\texttt{hep-th/9905104}}.

\bibitem{GM}
  S.~S.~Gubser and I.~Mitra,
  \textit{``Double trace operators and one loop vacuum energy in AdS/CFT''},
  \textsf{Phys.\ Rev.\ D {\bf 67}, 064018 (2003)},
  \href{http://arXiv.org/abs/hep-th/0210093}{\texttt{hep-th/0210093}}.

\bibitem{GK}
  S.~S.~Gubser and I.~R.~Klebanov,
  \textit{``A Universal result on central charges in the presence of double trace deformations''},
  \textsf{Nucl.\ Phys.\ B {\bf 656}, 23 (2003)},
  \href{http://arXiv.org/abs/hep-th/0212138}{\texttt{hep-th/0212138}}.


\bibitem{Hartman:2006dy}
  T.~Hartman and L.~Rastelli,
  \textit{``Double-trace deformations, mixed boundary conditions and functional determinants in AdS/CFT''},
  \textsf{JHEP {\bf 0801}, 019 (2008)},
  \href{http://arXiv.org/abs/hep-th/0602106}{\texttt{hep-th/0602106}}.


\bibitem{Herzog:2019bom}
  C.~P.~Herzog and I.~Shamir,
  \textit{``On Marginal Operators in Boundary Conformal Field Theory''}
  \href{http://arXiv.org/abs/1906.11281}{\tt arXiv:1906.11281}.

\bibitem{Polchinski:2011im}
  J.~Polchinski and J.~Sully,
  \textit{``Wilson Loop Renormalization Group Flows''},
  \textsf{JHEP~{\bf 1110},~059~(2011)},
  \href{http://arXiv.org/abs/1104.5077}{\tt arXiv:1104.5077}.

\bibitem{Drukker:2009hy}
  N.~Drukker and D.~Trancanelli,
  \textit{``A Supermatrix model for N=6 super Chern-Simons-matter theory''},
  \textsf{JHEP~{\bf 1002},~058~(2010)},
  \href{http://arxiv.org/abs/0912.3006}{\tt arXiv:0912.3006}.


\bibitem{Drukker:2008zx}
  N.~Drukker, J.~Plefka and D.~Young,
 \textit{``Wilson loops in 3-dimensional N=6 supersymmetric Chern-Simons Theory and their string theory duals''},
  \textsf{JHEP {\bf 0811} (2008) 019}
  \href{http://arxiv.org/abs/0809.2787}{\tt arXiv:0809.2787}.

\bibitem{Chen:2008bp}
  B.~Chen and J.~B.~Wu,
 \textit{``Supersymmetric Wilson Loops in N=6 Super Chern-Simons-matter theory''},
  \textsf{Nucl.\ Phys.\ B {\bf 825}, 38 (2010)}
  doi:10.1016/j.nuclphysb.2009.09.015
  \href{http://arxiv.org/abs/0809.2863}{\tt arXiv:0809.2863}.


  \bibitem{Rey:2008bh}
  S.~J.~Rey, T.~Suyama and S.~Yamaguchi,
  \textit{``Wilson Loops in Superconformal Chern-Simons Theory and Fundamental Strings in Anti-de Sitter Supergravity Dual''},
  \textsf{JHEP {\bf 0903} (2009) 127}
  \href{http://arxiv.org/abs/0809.3786}{\tt arXiv:0809.3786}.

\bibitem{Drukker:2019bev}
  N.~Drukker {\it et al.},
  \textit{``Roadmap on Wilson loops in 3d Chern-Simons-matter theories''},
  \href{http://arXiv.org/abs/1910.00588}{\tt arXiv:1910.00588}.

\bibitem{Lewkowycz:2013laa}
  A.~Lewkowycz and J.~Maldacena,
 \textit{``Exact results for the entanglement entropy and the energy radiated by a quark''}
 \textsf{JHEP {\bf 1405} (2014) 025}
  \href{http://arxiv.org/abs/1312.5682}{\tt arXiv:1312.5682}.

\bibitem{Ouyang:2015iza}
   H.~Ouyang, J.~B.~Wu and J.~j.~Zhang,
   \textit{``Novel BPS Wilson loops in three-dimensional quiver Chern-Simons-matter theories''},
   \textsf{ Phys.\ Lett.\ B {\bf 753}, 215 (2016)},
   \href{http://arXiv.org/abs/1510.05475}{\tt arXiv:1510.05475}.

\bibitem{Ouyang:2015bmy}
   H.~Ouyang, J.~B.~Wu and J.~j.~Zhang,
   \textit{``Construction and classification of novel BPS Wilson loops in quiver Chern-Simons matter theories''},
   \textsf{Nucl.\ Phys.\ B {\bf 910}, 496 (2016)},
   \href{http://arXiv.org/abs/1511.02967}{\tt arXiv:1511.02967}.

\bibitem{Lee:2010hk}
  K.~M.~Lee and S.~Lee,
  \textit{``1/2-BPS Wilson Loops and Vortices in ABJM Model''},
  \textsf{JHEP {\bf 1009} (2010) 004}
  \href{http://arxiv.org/abs/1006.5589}{\tt arXiv:1006.5589}.


\bibitem{Cardinali:2012ru}
  V.~Cardinali, L.~Griguolo, G.~Martelloni and D.~Seminara,
  \textit{``New supersymmetric Wilson loops in ABJ(M) theories''},
  \textsf{Phys.~Lett.~B~718, 615 (2012)},
  \href{http://arXiv.org/abs/1209.4032}{\tt arXiv:1209.4032}.


\bibitem{Correa:2014aga}
  D.~H.~Correa, J.~Aguilera-Damia and G.~A.~Silva,
  \textit{``Strings in $AdS_4 \times \mathbb{CP}^{3}$ Wilson loops in ${\mathcal N}=6$ super Chern-Simons-matter and bremsstrahlung functions''},
  \textsf{JHEP {\bf 1406} (2014) 139},
  \href{http://arXiv.org/abs/1405.1396}{\tt arXiv:1405.1396}.

\bibitem{Beccaria:2017rbe}
  M.~Beccaria, S.~Giombi and A.~Tseytlin,
  \textit{``Non-supersymmetric Wilson loop in $ \mathcal{N} $ = 4 SYM and defect 1d CFT''},
  \textsf{JHEP {\bf 1803}, 131 (2018)}
 \href{http://arXiv.org/abs/1712.06874}{\tt arXiv:1712.06874}.

\bibitem{cvet}
  M.~Cvetic, H.~Lu, C.~N.~Pope and K.~S.~Stelle,
  \textit{``T duality in the Green-Schwarz formalism, and the massless / massive IIA duality map''},
  \textsf{Nucl.\ Phys.\ B {\bf 573}, 149 (2000)},
 {\tt hep-th/9907202}.

\bibitem{Mart}
  L.~Martucci, J.~Rosseel, D.~Van den Bleeken and A.~Van Proeyen,
 \textit{``Dirac actions for D-branes on backgrounds with fluxes'',}
 \textsf{Class.\ Quant.\ Grav.\  {\bf 22}, 2745 (2005)}, \href{https://arxiv.org/abs/hep-th/0504041}
  {\tt hep-th/0504041}.



\bibitem{KK}
  H.~Kim, N.~Kim and J.~Hun Lee,
  \textit{``One-loop corrections to holographic Wilson loop in AdS4xCP3''},
  \textsf{J.\ Korean Phys.\ Soc.\  {\bf 61}, 713 (2012)},
  \href{http://arXiv.org/abs/1203.6343}{\tt arXiv:1203.6343}.

\bibitem{FLS}
  J.~Aguilera-Damia, A.~Faraggi, L.~A.~Pando Zayas, V.~Rathee and G.~A.~Silva,
  \textit{Toward Precision Holography in Type IIA with Wilson Loops,''}
  \textsf{JHEP {\bf 1808}, 044 (2018)},
  \href{http://arXiv.org/abs/1805.00859}{\tt arXiv:1805.00859 [hep-th].}.

\bibitem{Amsel:2008iz}
  A.~J.~Amsel and D.~Marolf,
  \textit{``Supersymmetric Multi-trace Boundary Conditions in AdS''},
  \textsf{Class.\ Quant.\ Grav.\  {\bf 26}, 025010 (2009)},
 \href{http://arXiv.org/abs/0808.2184}{\tt arXiv:0808.2184}.



\bibitem{Sakai:1984vm}
  N.~Sakai and Y.~Tanii,
  \textit{``Supersymmetry in Two-dimensional Anti-de Sitter Space''},
  \textsf{Nucl.\ Phys.\ B {\bf 258}, 661 (1985)}.

\bibitem{Drukker:2000ep}
  N.~Drukker, D.~J.~Gross and A.~A.~Tseytlin,
  \textit{``Green-Schwarz string in AdS(5) x S**5: Semiclassical partition function''},
  \textsf{JHEP {\bf 0004}, 021 (2000)},
  \href{http://arXiv.org/abs/hep-th/0001204}{\tt hep-th/0001204}.

\bibitem{Minces:1999eg}
  P.~Minces and V.~O.~Rivelles,
  \textit{``Scalar field theory in the AdS / CFT correspondence revisited,''}
 \textsf{ Nucl.\ Phys.\ B {\bf 572}, 651 (2000)} \href{http://arXiv.org/abs/hep-th/9907079},
  {\tt hep-th/9907079}.

\bibitem{Henneaux:1998ch}
  M.~Henneaux,
  \textit{``Boundary terms in the AdS / CFT correspondence for spinor fields''},
  \href{http://arXiv.org/abs/hep-th/9902137} {\texttt{hep-th/9902137}}.


\bibitem{Gubser:1998bc}
  S.~S.~Gubser, I.~R.~Klebanov and A.~M.~Polyakov,
  \textit{``Gauge theory correlators from noncritical string theory''},
  \textsf{Phys.\ Lett.\ B {\bf 428} (1998) 105},
  \href{http://arXiv.org/abs/hep-th/9802109}{\texttt{hep-th/9802109}}.

\bibitem{Witten:1998qj}
  E.~Witten,
  \textit{``Anti-de Sitter space and holography''}
  \textsf{Adv.\ Theor.\ Math.\ Phys.\  {\bf 2}, 253 (1998)},
  \href{http://arXiv.org/abs/hep-th/9802150}{\texttt{hep-th/9802150}}.

\bibitem{Freedman:1998tz}
  D.~Z.~Freedman, S.~D.~Mathur, A.~Matusis and L.~Rastelli,
  \textit{``Correlation functions in the CFT(d) / AdS(d+1) correspondence''},
  \textit{Nucl.\ Phys.\ B {\bf 546} (1999) 96},
  \href{http://arXiv.org/abs/hep-th/9804058}{\texttt{hep-th/9804058}}.


\bibitem{Sakai:1984nc}
  N.~Sakai and Y.~Tanii,
  \textit{``Supersymmetry and Vacuum Energy in Anti-de Sitter Space''},
  \textsf{Phys.\ Lett.\  {\bf 146B}, 38 (1984)}.


\bibitem{Camporesi:1992wn}
  R.~Camporesi and A.~Higuchi,
  \textit{``Stress energy tensors in anti-de Sitter space-time''},
  \textsf{Phys.\ Rev.\ D {\bf 45} (1992) 3591}.






\bibitem{Aguilera-Damia:2014bqa}
  J.~Aguilera-Damia, D.~H.~Correa and G.~A.~Silva,
  \textit{``Semiclassical partition function for strings dual to Wilson loops with small cusps in ABJM''},
  \textsf{JHEP {\bf 1503}, 002 (2015)},
  \href{http://arXiv.org/abs/1412.4084}{\tt arXiv:1412.4084}.




\bibitem{Marolf:2007in}
  D.~Marolf and S.~F.~Ross,
  \textit{``Reversing renormalization-group flows with AdS/CFT''},
  \textsf{JHEP {\bf 0805}, 055 (2008)},
  \href{http://arXiv.org/abs/0705.4642}{\tt arXiv:0705.4642}.


\bibitem{Compere:2008us}
  G.~Compere and D.~Marolf,
  \textit{``Setting the boundary free in AdS/CFT''},
  \textsf{Class.\ Quant.\ Grav.\  {\bf 25}, 195014 (2008)},
  \href{http://arXiv.org/abs/0805.1902}{\tt arXiv:0805.1902}.


\bibitem{Beccaria:2019dws}
  M.~Beccaria, S.~Giombi and A.~A.~Tseytlin,
  \textit{``Correlators on non-supersymmetric Wilson line in $ \mathcal{N}=4 $ SYM and AdS$_{2}$/CFT$_{1}$''},
  \textsf{JHEP {\bf 1905}, 122 (2019)},
  \href{http://arXiv.org/abs/1903.04365}{\tt arXiv:1903.04365}.


\bibitem{Bianchi:2017ozk}
  L.~Bianchi, L.~Griguolo, M.~Preti and D.~Seminara,
  \textit{``Wilson lines as superconformal defects in ABJM theory: a formula for the emitted radiation''},
  \textsf{JHEP {\bf 1710} (2017) 050},
  \href{http://arXiv.org/abs/1706.06590}{\tt arXiv:1706.06590}.

  \bibitem{Bianchi:2018scb}
  L.~Bianchi, M.~Preti and E.~Vescovi,
  \textit{``Exact Bremsstrahlung functions in ABJM theory''},
  \textsf{JHEP {\bf 1807} (2018) 060},
  \href{http://arXiv.org/abs/1802.07726}{\tt arXiv:1802.07726}.


\bibitem{Correa:2018fgz}
  D.~Correa, M.~Leoni and S.~Luque,
  \textit{``Spin chain integrability in non-supersymmetric Wilson loops''}
  \textsf{JHEP {\bf 1812}, 050 (2018)},
  \href{http://arXiv.org/abs/1810.04643}{\tt arXiv:1810.04643}.

\bibitem{Mauri:2017whf}
  A.~Mauri, S.~Penati and J.~j.~Zhang,
  \textit{``New BPS Wilson loops in $ \mathcal{N}=4 $ circular quiver Chern-Simons-matter theories''},
  \textit{JHEP {\bf 1711} (2017) 174},
  \href{http://arXiv.org/abs/1709.03972}{\tt arXiv:1709.03972}.


\bibitem{Fu:2016vas}
  W.~Fu, D.~Gaiotto, J.~Maldacena and S.~Sachdev,
  \textit{``Supersymmetric Sachdev-Ye-Kitaev models''}
  \textsf{Phys.\ Rev.\ D {\bf 95} (2017) no.2,  026009
   Addendum: [Phys.\ Rev.\ D {\bf 95} (2017) no.6,  069904]}
    \href{http://arXiv.org/abs/hep-th/1610.08917}{\texttt{hep-th/1610.08917}}.


\bibitem{LPT}
  H.~Lu, C.~N.~Pope and P.~K.~Townsend,
  \textit{``Domain walls from anti-de Sitter space-time''},
  \textsf{Phys.\ Lett.\ B {\bf 391}, 39 (1997)},
  \href{http://arXiv.org/abs/hep-th/9607164}{\texttt{hep-th/9607164}}.



\end{thebibliography}

\end{document}